\documentclass[preprint2]{aastex6}
\usepackage[utf8]{inputenc}
\usepackage{color}
\usepackage{enumitem}
\usepackage[fleqn]{amsmath}
\usepackage{bm}
\usepackage{url}
\usepackage{natbib}
\usepackage{csvsimple}

\makeatletter
\def\env@matrix{\hskip -\arraycolsep 
  \let\@ifnextchar\new@ifnextchar
  \array{*{\c@MaxMatrixCols}c}}
\makeatother

\shorttitle{Non-Transiting Hot Jupiters with Phase Curves}
\shortauthors{Millholland \& Laughlin}

\begin{document}

\title{Supervised Learning Detection of Sixty Non-Transiting Hot Jupiter Candidates}
\author{Sarah Millholland and Gregory Laughlin}
\affil{Department of Astronomy, Yale University, New Haven, CT, 06511, USA}
\email{sarah.millholland@yale.edu}

\begin{abstract}
The optical, full-phase photometric variations of a short-period planet provide a unique view of the planet's atmospheric composition and dynamics. The number of planets with optical phase curve detections, however, is currently too small to study them as an aggregate population, motivating an extension of the search to non-transiting planets. Here we present an algorithm for the detection of non-transiting, short-period giant planets in the \textit{Kepler} field. The procedure uses the phase curves themselves as evidence for the planets' existence. We employ a supervised learning algorithm to recognize the salient time-dependent properties of synthetic phase curves; we then search for detections of signals that match these properties. After demonstrating the algorithm's capabilities, we classify 142,630 FGK \textit{Kepler} stars without confirmed planets or KOIs and, for each one, assign a probability of a phase curve of a non-transiting planet being present. We identify 60 high-probability non-transiting hot Jupiter candidates. We also derive constraints on the candidates' albedos and offsets of the phase curve maxima. These targets are strong candidates for follow-up radial velocity confirmation and characterization. Once confirmed, the atmospheric information content in the phase curves may be studied in yet greater detail.

\end{abstract}

\section{Introduction}

While the \textit{Kepler} prime mission is best known for the detection of thousands of transiting extrasolar planets, notably including some that are Earth-sized and smaller \citep[e.g.][]{2013ApJS..204...24B}, its long time baseline and high photometric precision allowed for the detection of other small-amplitude signals, such as out-of-transit photometric variations of short-period planets \citep{2009Sci...325..709B}. Long before \textit{Kepler} was launched, the existence and detectability of these optical reflected light phase curves were predicted by \cite{2000ApJ...540..504S} and \cite{2003ApJ...595..429J}.

Optical phase curves, along with their infrared counterparts, provide a spatially-integrated, time-dependent view of the planet's atmosphere. They yield information about the planet's atmospheric composition \citep{2006ApJ...646.1241R, 2013ApJ...776L..25D}, day-night temperature contrasts and heat redistribution \citep{2007Natur.447..183K, 2009ApJ...699..564S, 2011ApJ...729...54C}, cloud existence and reflectivity \citep{2013ApJ...776L..25D, 2015AJ....150..112S}, atmospheric weather variability \citep{2016NatAs...1E...4A}, and magnetic field strength \citep{2017arXiv170406271R}. A recent review of phase curve methodology and scientific findings is provided by \cite{2017arXiv170300496S}.

A planet's optical phase curve is the superposition of four independent components: reflected star light, thermal emission, ellipsoidal variations, and Doppler beaming/boosting \citep{2011MNRAS.415.3921F}. Ellipsoidal variations are brightness changes resulting from tidal deformation to the star \citep{1985ApJ...295..143M}, while Doppler beaming is a relativistic effect that causes an increase or decrease in light as the star approaches or recedes from the observer \citep{1974A&A....30..135H, 1979rpa..book.....R}. Ellipsoidal variations and Doppler beaming are both sensitive to the planet's mass, which thereby permits photometric mass constraints \citep[e.g.][]{2016A&A...592A..32L}. For an average hot Jupiter ($P \sim 3$ days, $R_p \sim 1.3 \ R_{\mathrm{Jup}}$, $M_p \sim M_{\mathrm{Jup}}$) orbiting a Sun-like star, the reflected light component is typically larger than the other components by about an order of magnitude.

There have been numerous detections of optical phase curves of transiting \textit{Kepler} planets \citep[e.g.][]{2010ApJ...713L.145W, 2011AJ....142..195S, 2012ApJ...761...53B, 2012A&A...541A..56M, 2014A&A...562A.109L}. In addition, comprehensive searches for \textit{Kepler} planetary phase curves and secondary eclipses have also been published \citep{2012AJ....143...39C, 2013ApJ...772...51E, 2015PASP..127.1113A, 2015ApJ...804..150E, 2016A&A...592A..32L}. These systematic searches have established several interesting findings, including generally low (but occasionally quite large) hot Jupiter albedos \citep{2011ApJ...735L..12D} and repeated detections of shifts in the maximum of the phase curve away from the sub-stellar point \citep{2013ApJ...771...26F, 2015ApJ...800...73F, 2015AJ....150..112S, 2015ApJ...802...51H}.

Though it is substantial, the sample size ($\sim$ 15) of \textit{Kepler} transiting planets with detectable photometric variations is just short of the count necessary to adequately understand these atmospheric properties in a statistical or populational sense. It is therefore worthwhile to increase the number of planets with well-characterized optical phase curves by expanding the search to non-transiting planets and using the phase curve signal itself as a detection mechanism. Given their large expected radial velocity half-amplitudes, candidates identified in this manner can subsequently be confirmed with Doppler velocity measurements.

The prospects of detecting non-transiting planets from their photometric variations alone has been frequently discussed in the literature. The BEaming, Ellipsoidal, and Reflection/heating (BEER) model \citep{2011MNRAS.415.3921F} was developed in part as a technique to discover non-transiting planets. To date, the BEER model has been very successful in the detection of non-transiting stellar binary companions \citep{2012ApJ...746..185F} and the characterization of transiting planets with detectable phase curve variations \citep{2013ApJ...771...26F, 2015ApJ...800...73F}.

 \cite{2014ApJ...795..112P} proposed using Bayesian model selection to test a light curve for the presence of significant reflected light, thermal emission, ellipsoidal variation, and/or Doppler beaming phase curve components. \cite{2016ApJ...823L...7M} considered the prospects of discovering non-transiting hot Jupiters in \textit{Kepler} systems containing known transiting planets with the coupled detection of an optical phase curve and astrometric (stellar wobble-induced) transit timing variations.
 
 In all of these techniques, the primary challenge 
 is that the ellipsoidal and Doppler beaming components are typically significantly smaller than the reflected light component for Jupiter-sized planets. The photometric signature is therefore mostly sinusoidal and difficult to distinguish from stellar or instrumental variability. These prior studies, however, have uniformly considered the phase-folded and time-averaged phase curve signal. When the signal is folded and binned, valuable time-dependent information is lost. This normally discarded information can be helpful in distinguishing planetary signals from stellar ones.

In this paper, we present a systematic procedure to detect non-transiting, short-period giant planets from their phase curves by exploiting the phase curves' time-dependent properties.
The algorithm is rooted in two key ideas:  \\
\begin{enumerate}
\item A planetary phase curve is more temporally consistent in its amplitude and phase than most other stellar or instrumental light curve variability. 
\item The properties of phase curves of known transiting \textit{Kepler} planets and synthetically generated datasets can be used as a training set on which novel light curves can be compared and classified using supervised learning.
\end{enumerate}


This paper is organized as follows. In \S2, we outline the pipeline we developed for detecting candidate phase curves in \textit{Kepler} light curves. In \S3, we inject synthetic phase curves into \textit{Kepler} light curves and demonstrate the pipeline's efficiency in recovering them. In \S4, we train a logistic regression algorithm to distinguish the properties of planetary phase curves from other signals. We test its classification capabilities on synthetic datasets and the set of \textit{Kepler} transiting hot Jupiters. In \S5, we employ the supervised learning algorithm to search for phase curves around \textit{Kepler} FGK stars without known planets or planet candidates. In \S6, we establish a catalog of candidate planets with phase curves and examine trends in the planets' albedos and phase curve maximum offsets. We discussion additional properties of the candidates in \S7 and conclude in \S8.

\section{An Automated Detection Pipeline}\label{Pipeline}

To detect non-transiting hot Jupiters in the \textit{Kepler} field, we first construct an algorithm for identifying the single best candidate phase curve within a given light curve. The process consists of three distinct steps:
\begin{enumerate}
\item{Search for the periods of potentially significant, approximately sinusoidal signals in the data by calculating Lomb-Scargle periodograms of initially detrended light curves.}
\item{At each candidate period, find the best-fit parameters of a phase curve by simultaneously fitting for the phase curve and stellar components.}
\item{Quantify the autocorrelations in the residuals of the fits.} 
\end{enumerate}

The light curves we consider here are the PDC (pre-search data conditioning) photometry \citep{2012PASP..124.1000S, 2012PASP..124..985S, 2014PASP..126..100S} provided by the \textit{Kepler} Science Center and publicly available at Mikulski Archive for Space Telescopes (MAST)\footnote{\href{https://archive.stsci.edu/kepler/}{https://archive.stsci.edu/kepler/}}. The PDC light curves have undergone a removal of systematic errors common to all light curves.

\subsection{Step 1- Identifying candidate periods}
\label{pipeline step 1}

The first goal is to identify the periods of significant sinusoidal signals in the light curve. To do this, we must first damp the effects of stellar variability or residual instrumental systematics that could contaminate a periodogram. We model our detrending algorithm on the \textit{kepflatten} routine in the PyKE software package\footnote{\href{https://keplerscience.arc.nasa.gov/PyKE.shtml}{https://keplerscience.arc.nasa.gov/PyKE.shtml}} \citep{2012ascl.soft08004S}. 

We detrend the light curve by fitting it with the mean of polynomials that span a sliding window. In detail, we first split the light curve into sections of length $s$. Then, over a window of size $w = Ns$ (with $N$ an integer), we fit a third degree polynomial. We slide the window through all time sections, stepping in units of $s$. Within each section, we then calculate the mean of the $N$ polynomials that were fit in that section, resulting in a smooth fitted curve. We then detrend the light curve data by dividing by the fit. At this stage, we calculate outliers that are deviant by more than $7\sigma$, remove them, and iterate the process. 

We perform two detrending calculations with different window sizes in an attempt to account for stellar variability at a variety of timescales. A step size of $s=2$ days is used in each calculation, but the short-timescale detrending uses $w=6$ days, while the long-timescale detrending uses $w=12$ days. We aim to detect phase curves with orbital periods up to 5 days (though in reality, most detections will be less than 3 days). It is therefore critical that $w$ be large enough to avoid fitting out the phase curve signals we wish to detect. This explains our choice of $w=6$ for the short-timescale detrending.

On each of the two detrended curves associated with a given light curve, we generate a Lomb-Scargle periodogram \citep{1976ApSS..39..447L, 1982ApJ...263..835S} within a 1-5 day range. Hence, we assume that we will only be capable of detecting phase curves within this period range. Although there are several known hot Jupiters with sub-day periods \citep[e.g. WASP-18b,][]{2009Natur.460.1098H}, we found that extending the periodogram much below 1 day leads to significantly more noise and poorer overall results.

\begin{figure}
\epsscale{1.27}
\plotone{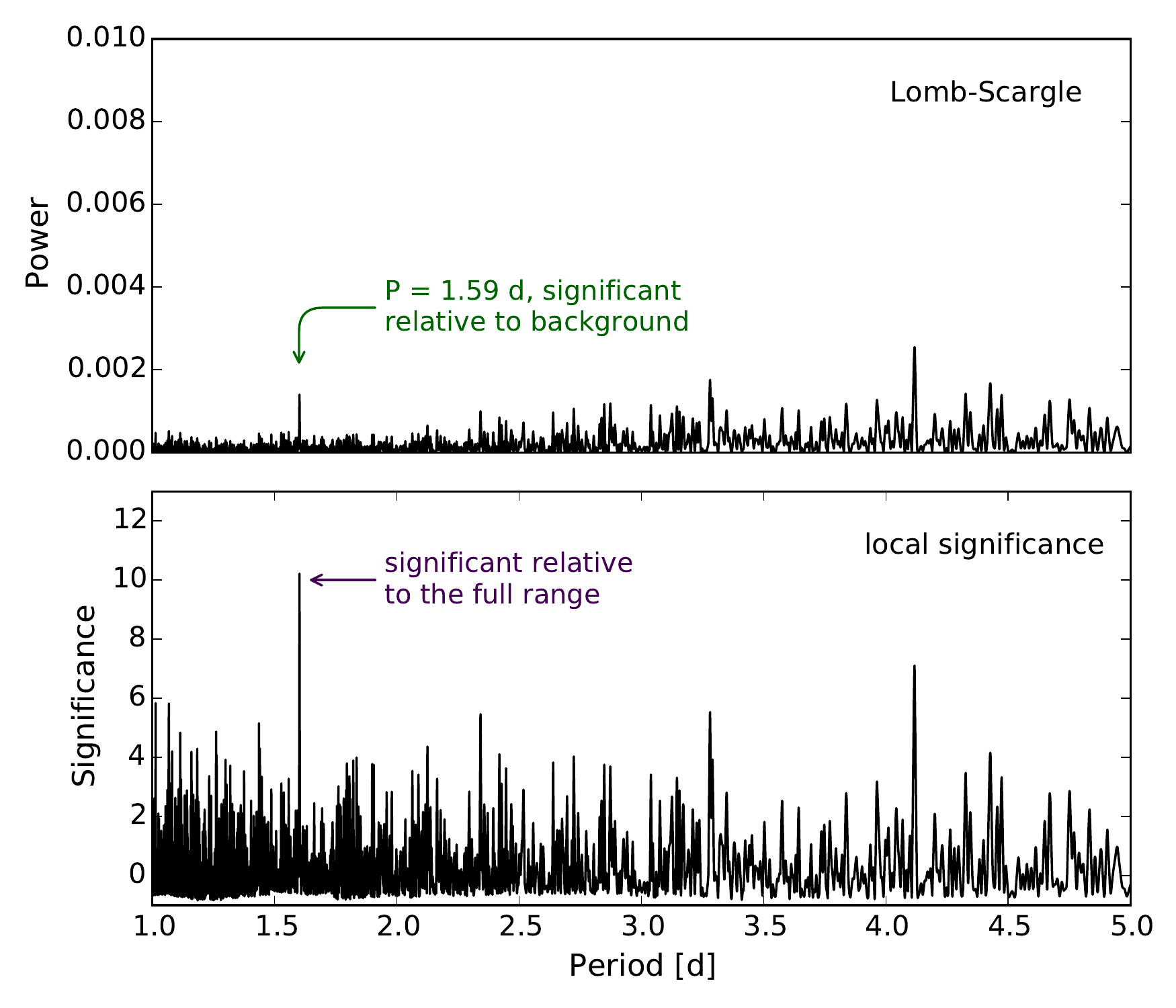}
\caption{The Lomb-Scargle periodogram and local significance periodogram of the light curve of the transiting hot Jupiter, Kepler-686b (KIC 3935914). Transits and secondary eclipses were removed before computation. In this example, the peak associated with the true period is not the highest in the Lomb-Scargle periodogram, but it is the highest in the local significance periodogram.}
\label{periodograms}
\end{figure}

We also generate a ``local significance'' periodogram. At each period in the spectrum, we calculate the number of standard deviations of the periodogram power over the median power within a centered 0.5 day window. The purpose of this additional step is that the peak in a periodogram associated with a true phase curve is sometimes highly significant with respect to nearby powers but not with respect to the entire 1-5 day range. As an illustrative example, Figure~\ref{periodograms} displays the Lomb-Scargle periodogram and local significance periodogram for the $R \sim 1.08 \ R_{\mathrm{Jup}}$, $P \sim 1.595$ day, transiting hot Jupiter, Kepler-686b. We removed the transits and secondary eclipses in the light curve before the computation.

We extract the highest peak in each of the four periodograms -- one Lomb-Scargle and one local significance periodogram for each of the two detrending calculations. We find the unique periods (defined as separate by more than 0.01 days) among these, yielding a collection of 1-4 peak periods. These period(s) constitute candidate periods for a planetary phase curve. The candidate signals are analyzed in more detail in the next section. 

\subsection{Step 2- Phase curve fitting}
\label{phase curve fitting}

The initial detrending calculations described above are sufficient for identifying the periods of significant sinusoidal signals in the data. To robustly characterize the candidate phase curves, however, it is necessary to model the planetary signal and stellar variability simultaneously, rather than attempting to remove the stellar signal \textit{a priori}. To this end, we use the following steps to find the best-fit sinusoidal signal for each of the candidate period(s) identified.

\begin{figure}
\epsscale{1.2}
\plotone{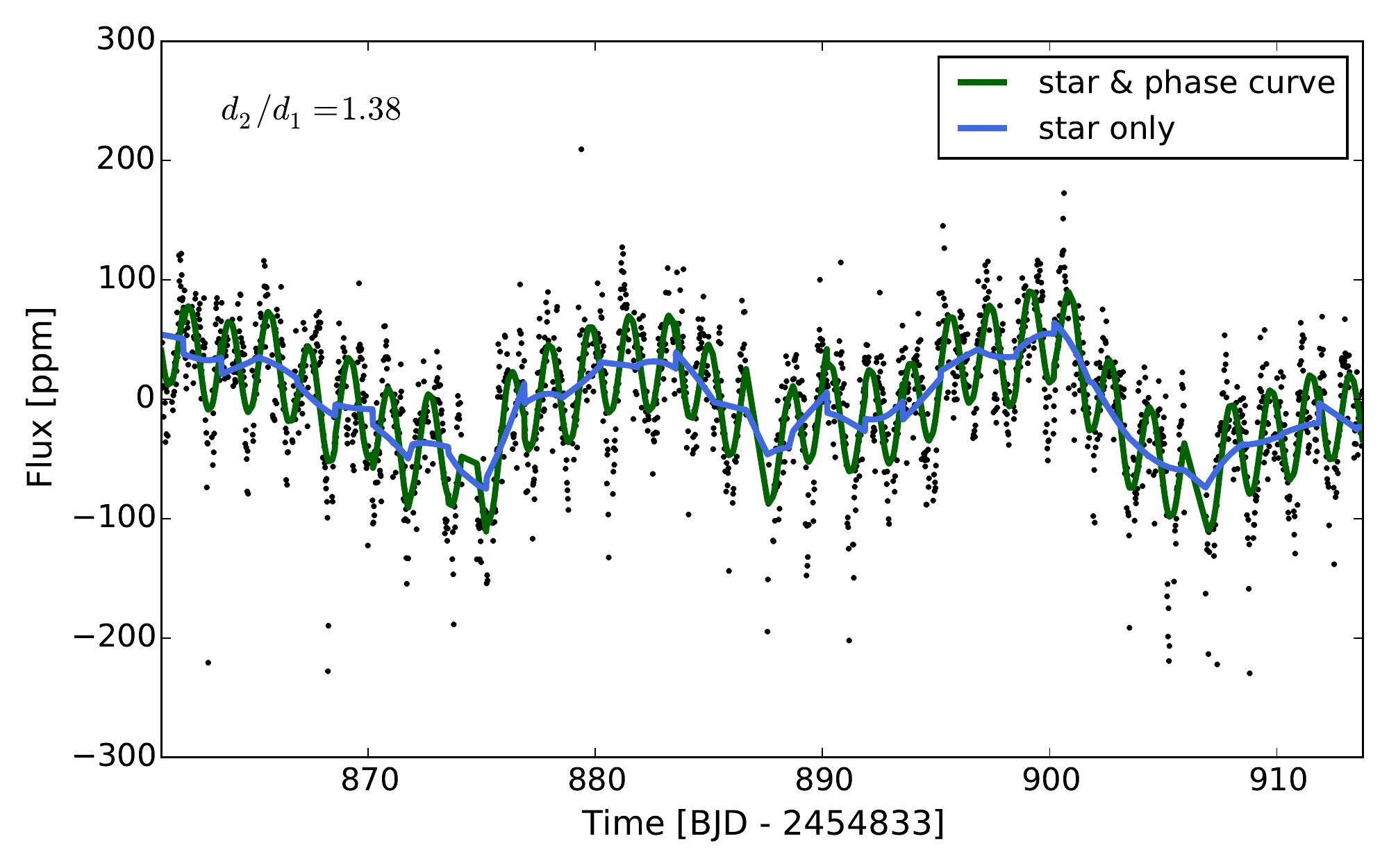}
\caption{The one- and two-component fits in a section of the light curve of Kepler-13b (KIC 9941662). The transits and secondary eclipses have been removed from the data. The black points are the PDC flux measurements. The green curve is the two-component, phase curve and stellar variability fit, while the blue curve is the stellar component only. The ratio of the Durbin-Watson statistics of the residuals of the two fits is displayed in the upper left.}
\label{phase curve fit}
\end{figure}

The best fit is found by chi-square minimization of a multiplicative, two-component (phase curve \& stellar variability) light curve model. The phase curve component is modeled by a fixed-period sinusoid with variable phase and amplitude. The stellar variability component, that is, the component remaining after the sinusoid is divided out, is fit using the sliding polynomial technique described in Section~\ref{pipeline step 1}. Here we use three degree polynomials, a step size $s = 2$ days, and window size $w = 6$ days. The best-fit is characterized by the amplitude and phase of the sinusoid that minimizes the residuals of the joint fit. For the purposes of later comparison, we also perform a one-component fit without the sinusoid. In other words, this just contains the sliding polynomial fit. An example of the one-component (star only) and two-component (star \& phase curve) fits are shown for Kepler-13b in Figure~\ref{phase curve fit}. Kepler-13b is a $R \sim 1.51 \ R_{\mathrm{Jup}}$, $P \sim 1.764 \ \mathrm{day}$, transiting hot Jupiter with a large amplitude phase curve. 

We note that techniques such as Gaussian Process (GP) regression or autoregressive integrated moving average (ARIMA) modeling might be better suited to capturing the stellar variability component of the light curve \citep[e.g.][]{2014MNRAS.443.2517H, 2015ApJ...800...46B, 2015IAUGA..2254945C}. We did successfully apply a GP regression to take the place of the stellar variability polynomial fit, but we found that any improvements in the fit did not outweigh the larger computational costs. 

\subsection{Step 3- Quantifying residual autocorrelations}
\label{residual autocorrelations}

Given the best-fit sinusoidal signal in the data for each of the 1-4 candidate period(s), our goal is to diagnose each signal's phase and amplitude coherence. This information will be used to select the best of the 1-4 candidate signals and, later in Section~\ref{Logistic Regression}, to recognize the time-dependent properties of true phase curves. 

The phase and amplitude coherence is best examined with the autocorrelation of the residuals of the fit. If the signal experiences large phase and amplitude variations, a sinusoidal fit will be inappropriate, and the residuals will be significantly non-Gaussian. On the other hand, the signal of a true phase curve will have less correlated residuals. We examine two diagnostics of the residual autocorrelation: the Durbin-Watson statistic \citep{doi:10.1093/biomet/37.3-4.409} and the Ljung-Box statistic \citep{doi:10.1093/biomet/65.2.297}. 

The Durbin-Watson test statistic for residuals $\{r_i\}$ is given by
\begin{equation}
\label{Durbin-Watson}
d = \frac{\sum\limits_{i =2}^{N}(r_i - r_{i-1})^2}{\sum\limits_{i=1}^{N}{r_i}^2}.
\end{equation}
In a Durbin-Watson test, $d$ is compared to critical values at selected significance levels. The test is used to investigate the null hypothesis that the residuals exhibit zero autocorrelation against the alternative hypothesis that the residuals are positively or negatively autocorrelated. Uncorrelated residuals have $d \sim 2$, while positively correlated residuals have $d \ll 2$. 

\begin{figure}
\epsscale{1.25}
\plotone{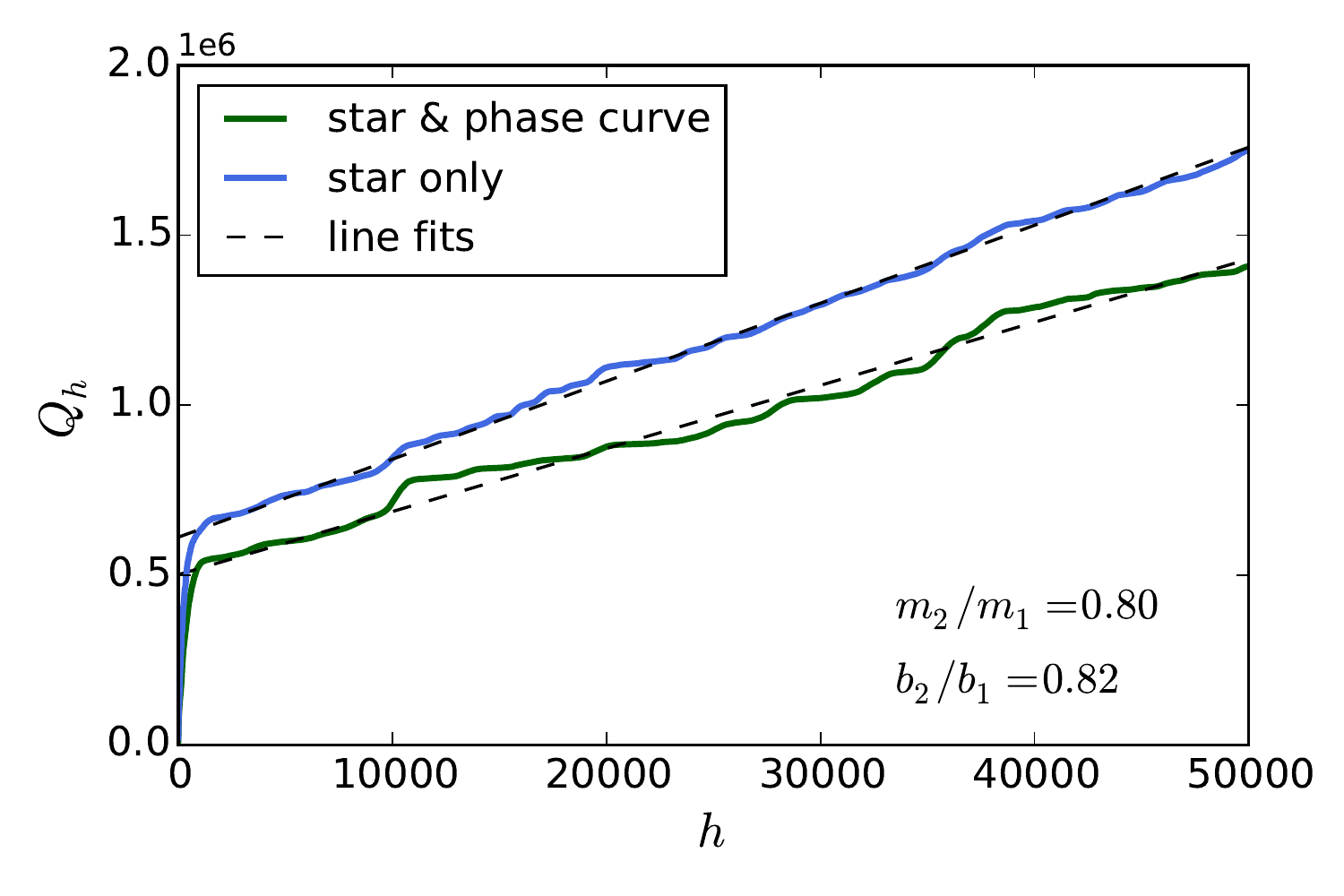}
\caption{The Ljung-Box statistic, $Q_h$, as a function of lag, $h$, for the residuals of the one-component and two-component light curve fits of Kepler-13b (KIC 9941662). A portion of the light curve fits were shown in Figure~\ref{phase curve fit}. The black dashed lines are linear fits. The ratios of the slopes and y-intercepts for the two lines are shown in the lower right.}
\label{Ljung_Box}
\end{figure}

The Ljung-Box statistic is given by
\begin{equation}
Q_h = N(N+2)\sum\limits_{j=1}^{h}\frac{{\hat{\rho_j}}^2}{N-j},
\end{equation}
where $\hat{\rho_j}$ is the autocorrelation of the residuals at lag $j$, $N$ is the number of data points, and $h$ is the number of lags being tested. If the residuals are independently distributed, $Q_h$ follows a chi-squared distribution with $h$ degrees of freedom, $Q_h \sim {\chi_h}^2$. The residuals show evidence for autocorrelation at significance level $\alpha$ if $Q_h > \chi_{1-\alpha, h}^2$. Rather than calculate the statistic at one lag, we calculate it at a variety of lags and plot $Q_h$ vs. $h$. For example, in Figure~\ref{Ljung_Box} we display the $Q_h$ vs. $h$ curves for the residuals of the one- and two-component fits of Kepler-13b. (A section of these fits were shown in Figure~\ref{phase curve fit}.) $Q_h$ exhibits a nearly linear dependence with the lag, with different slopes and y-intercepts between the two fits. Since the ``star only'' fit has more autocorrelated residuals, the slope and y-intercept of its $Q_h$ curve are larger.

Given that the two-component (phase curve \& stellar variability) fit described in Section~\ref{phase curve fitting} will rarely perfectly model a \textit{Kepler} light curve, there will always be some serial autocorrelation. Therefore to make the Durbin-Watson and Ljung-Box statistics useful, we do not perform the statistical tests directly. Rather, we use the ratio of the test statistics in the two-component fit to their values in the one-component fit. For the Durbin-Watson statistic and Ljung-Box slope and y-intercept, these ratios are denoted $d_2/d_1$, $m_2/m_1$, and $b_2/b_1$, respectively. The ratios for Kepler-13b are displayed in Figures~\ref{phase curve fit} and~\ref{Ljung_Box}.

As described in Section~\ref{phase curve fitting}, the one-component fit is the ``star only'' case, that is, the sliding polynomial fit to the data without a sinusoidal component. The ratio of the test statistics in the star \& phase curve vs. star only cases thus probes the degree to which the serial autocorrelation is improved or degraded by the inclusion of a sinusoidal signal in the data. Large signal-to-noise phase curves will exhibit a significant reduction in the autocorrelation when the two-component fit is performed, with $d_2/d_1 \gg 1$, $m_2/m_1 \ll 1$, $b_2/b_1 \ll 1$. Similarly, we also examine the degree to which the two-component fit is favored using the ratio of the $\chi^2$, denoted by ${\chi^2}_2/{\chi^2}_1$. Strongly favored phase curve fits will have ${\chi^2}_2/{\chi^2}_1 \ll 1$.

Returning to the description of the pipeline, at this stage the autocorrelation diagnostics have been calculated for the best-fit sinusoidal signals associated with each of the 1-4 peak periods. To choose a single best signal, we use the following minimization:
\begin{equation}
\min_{P \in \{P_1,...,P_4\}} \left\{\frac{{\chi^2}_2}{{\chi^2}_1} - \frac{d_2}{d_1} + \frac{m_2}{m_1}\right\}. \\
\end{equation}
We thus select the signal with the best fit and lowest autocorrelation. This step concludes the procedure for identifying candidate phase curves in \textit{Kepler} light curves.

\section{Synthetic Phase Curve Injection and Recovery}
\label{synthetic injection and recovery}

In Section~\ref{Pipeline}, we built a pipeline for the detection of a candidate phase curve signal in a \textit{Kepler} light curve. We now construct and inject synthetic phase curves into \textit{Kepler} light curves and demonstrate the pipeline's recovery efficiency. These synthetic phase curves will also serve as the training set for the logistic regression in Section~\ref{Logistic Regression}.

\subsection{Phase curve model} \label{phase curve model}

Optical phase curve models have been constructed by various authors, including \cite{2011MNRAS.415.3921F, 2012ApJ...761...53B, 2014ApJ...795..112P, 2015PASP..127.1113A} and \cite{2013ApJ...772...51E, 2015ApJ...804..150E}. Here we adopt the notation and model of \cite{2015ApJ...804..150E}. The light curve of a non-transiting planet is composed of three components: the flux coming from the planet (both reflection and thermal), $F_p$, and the Doppler beaming and ellipsoidal variation components from the star, $F_d$ and $F_e$. The composite light curve is given by 
\begin{equation}
F = F_p(\phi + \theta) + F_d(\phi) + F_e(\phi). 
\end{equation}
The curve is a function of the phase, $\phi$, which ranges from 0 to 1 and is given by 
\begin{equation}
\phi = \frac{t - T_{\mathrm{mid}}}{P},
\end{equation}
where $T_{\mathrm{mid}}$ is the time of inferior conjunction (or mid-transit for a transiting planet).

The planetary brightness is modeled as a Lambert sphere, described by 
\begin{equation}\label{planetary brightness}
F_p = A_p \frac{\sin z + (\pi - z) \cos z}{\pi} \, ,
\end{equation}
where $\cos z = - \sin i \cos(2\pi \left[ \phi + \theta\right])$.
The planetary brightness is approximately
\begin{equation}
\label{Ap equation}
A_p \approx F_{\mathrm{ecl}} = A_g \left(\frac{R_p}{a}\right)^2,
\end{equation}
and $\theta$ is a phase offset of the peak brightness from the sub-stellar point.

The Doppler beaming flux is given by 
\begin{equation}\label{Doppler beaming}
F_d = \left( \frac{2 \pi G}{P} \right)^{1/3} \frac{M_p \alpha_d \sin i}{c \ {M_{\star}}^{2/3}} \left( \frac{1+ e \cos\omega}{\sqrt{1-e^2}} \right) \sin(2\pi \phi).
\end{equation}
The constant $\alpha_d$ is the photon-weighted, bandpass-integrated beaming coefficient. 

The ellipsoidal variation component is, using the first three cosine harmonics,
\begin{multline}\label{ellipsoidal variation}
F_e = - \frac{M_p \alpha_2 \sin^2i}{M_{\star}} \biggl[ \left(\frac{a}{R_{\star}} \right)^{-3} \cos(4 \pi \phi) \ + \\ 
 \left(\frac{a}{R_{\star}} \right)^{-4} f_1 \cos(2 \pi \phi) + \left(\frac{a}{R_{\star}} \right)^{-4} f_2 \cos(6 \pi \phi)\biggr],
\end{multline}
where 
\begin{equation}
\begin{split}
f_1 &= 3 \alpha_1 \frac{5 \sin^2 i - 4}{\sin i} \\
f_2 &= 5 \alpha_1 \sin i
\end{split}
\end{equation}
and $\alpha_1$ and $\alpha_2$ are functions of the limb darkening and gravity darkening parameters \citep{2013ApJ...772...51E}.

\begin{figure}
\epsscale{1.2}
\plotone{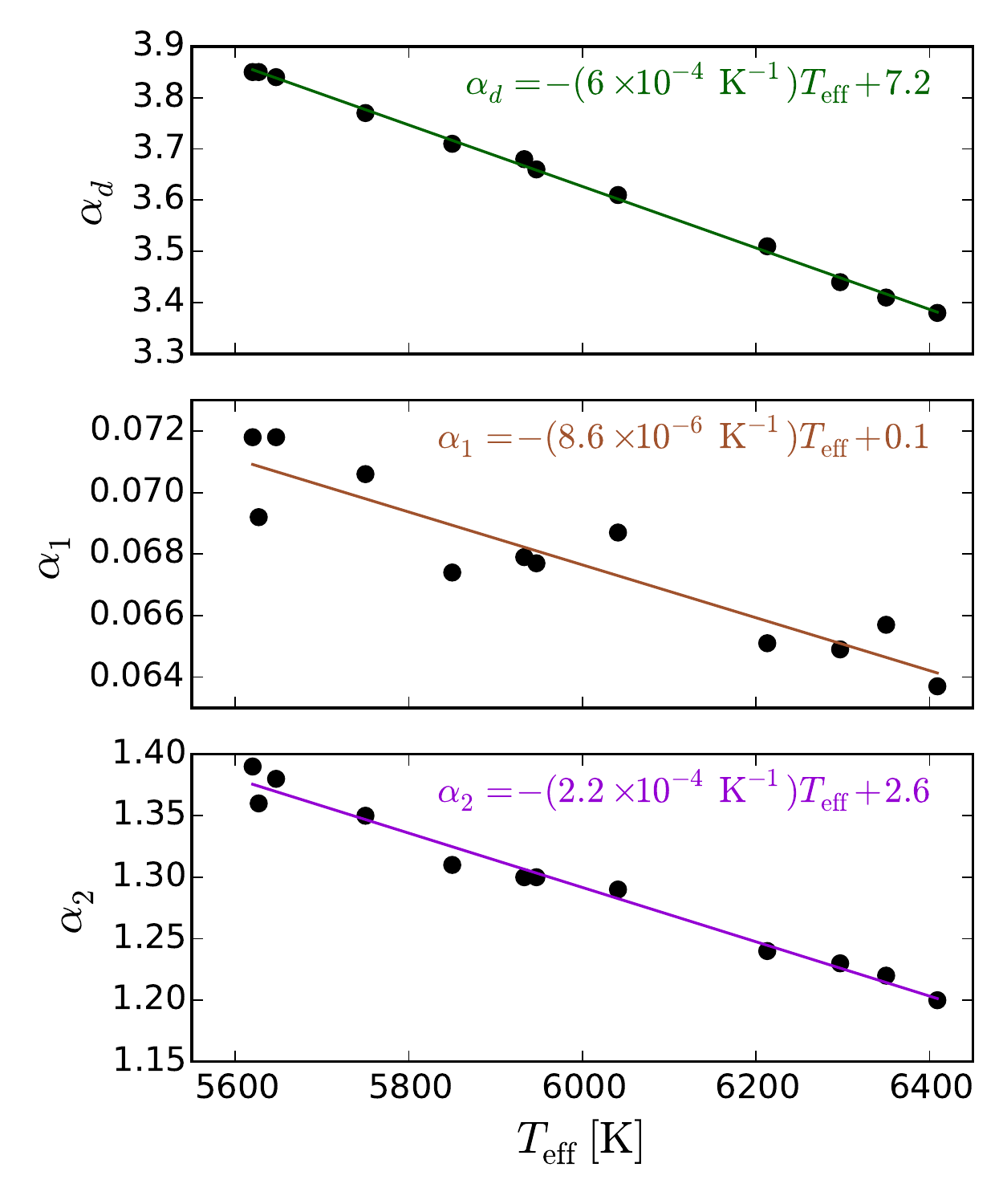}
\caption{The Doppler beaming constant, $\alpha_d$, and ellipsoidal variation constants, $\alpha_1$ and $\alpha_2$, versus $T_\mathrm{eff}$ for 12 stars in \cite{2015ApJ...804..150E}. The equations in the upper right corners correspond to the lines of best-fit. The quality of the fits illustrates that parametrization of these coefficients as simple functions of $T_\mathrm{eff}$ should be sufficient for our modeling purposes.}
\label{alpha constants}
\end{figure}

Though $\alpha_d$ is a function of the stellar spectrum and the \textit{Kepler} transmission function, and $\alpha_1$ and $\alpha_2$ are complicated functions of limb darkening parameters, they all exhibit nearly linear dependence with $T_\mathrm{eff}$ within the range of interest ($4000 \ \mathrm{K} < T_{\mathrm{eff}} < 7000 \ \mathrm{K}$). In Figure~\ref{alpha constants}, we plot the Doppler and ellipsoidal coefficients versus $T_\mathrm{eff}$ for 12 host-stars of planets with phase curves from \cite{2015ApJ...804..150E}. We also show best-fit lines. There is a clear linear relationship for each $\alpha$ coefficient with $T_{\mathrm{eff}}$. The trends agree well with an unpictured, additional star with $T_{\mathrm{eff}} = 4550 \ \mathrm{K}$.

Given the high quality of the fits, we simplify our phase curve model by letting $T_{\mathrm{eff}}$ be a parameter, with the $\alpha$ coefficients calculated using the best-fit lines pictured in Figure~\ref{alpha constants}. Considering equations~\ref{planetary brightness},~\ref{Doppler beaming}, and~\ref{ellipsoidal variation}, there are 9 parameters required to model a phase curve: $P, \ M_p, \ R_p, \ i, \ A_g, \ \theta, \ M_{\star}, \ R_{\star},  \ T_\mathrm{eff}$. We assume that $e \approx 0$ since planets with detectable phase-curves are typically hot Jupiters with low eccentricities.

\subsection{Light curve injection}
\label{light curve injection}

In order to inject simulated phase curves into \textit{Kepler} datasets, the most thorough procedure would be to insert the signal into pixel-level data, as has been done in measuring the transit signal recovery of the \textit{Kepler} pipeline \citep{2013ApJS..207...35C, 2015ApJ...810...95C, 2016ApJ...828...99C}. However, \cite{2013ApJS..207...35C} showed an extremely high fidelity in the preservation of individual injected transit signals from the pixel-level data through image calibration, aperture photometry, and pre-search data conditioning (PDC) systematics removal. Given this precedent, it should be appropriate to inject synthetic phase curves into the PDC light curve data, rather than the pixel-level data. 

We picked a random sample of 10,000 \textit{Kepler} FGK main sequence stars without confirmed planets or KOIs and extracted their PDC light curves. Here we define FGK dwarfs with simple cuts used by \cite{2016ApJ...828...99C}: $4000 \ K < T_{\mathrm{eff}} < 7000 \ K$, $\mathrm{log}\,g > 4.0$. For each target, we generated a single simulated phase curve using the nine physical quantities listed in Table 1. Here $\mathcal{U}$ denotes a uniform distribution. The stellar parameters, $\ M_{\star}, \ R_{\star}$, and $T_\mathrm{eff}$, are absent from this list. Rather than drawing these parameters randomly, we use the values of the target star. It is important to note that the parameter space spanned by the distributions in Table 1 need not reflect the true distribution of planets; rather, they must encompass all plausible phase curve amplitudes and morphologies capable of being observed.

\begin{table}[!h]
\caption{Parameter Distributions of Synthetic Phase Curve Injections} \label{synthetic injection parameters} 
\begin{center}
\begin{tabular}{ c | c} 
 \hline
 \hline
 Parameter & distribution \\
 \hline
period, $P$ [days] & $P \sim \mathcal{U}[1, 5]$ \\
planet mass, $M_{P}$ [$M_{\mathrm{Jup}}$] & $M_{P} \sim \mathcal{U}[0.25, 2.5]$ \\
planet radius, $R_{P}$ [$R_{\mathrm{Jup}}$] & $R_{P} \sim \mathcal{U}[0.7, 1.8]$ \\ 
inclination, $i$ [$^{\circ}$] & $i \sim \mathcal{U}[45, 90]$ \\
mean geometric albedo, $\left<A_g\right>$ & $\left<A_g\right> \sim \mathcal{U}[0.03, 0.35]$ \\
albedo covariance amplitude, ${h_A}^2$ & ${h_A}^2 \sim \mathcal{U}[0, 0.015]$ \\
mean peak offset, $\left<\theta\right>$ & $\left<\theta\right> \sim \mathcal{U}[-0.1, 0.05]$ \\
$\theta$ covariance amplitude, ${h_{\theta}}^2$ & ${h_{\theta}}^2 \sim \mathcal{U}\left[0,0.0025/{\left<\theta\right>}^2\right]$ \\
coherence timescale, $\tau$ [days] & $\tau \sim  \mathcal{U}[10, 50]$ \\
 \hline
\end{tabular}
\end{center}
\end{table}

There are three additional parameters in Table 1 that have not yet been introduced: ${h_A}^2$, ${h_{\theta}}^2$, and $\tau$. These are related to modulations we impose on the planet's albedo and phase offset. This variability is motivated by the assumption that planets experience surface feature evolution and atmospheric flow that causes their spatially-dependent reflectivity to change slightly over time, which in turn influences the amplitudes and shapes of their observed phase curves. While the magnitude and timescale of this effect is unknown, a fractional albedo variation of up to $\sim30\%$ seems plausible \citep{2008ApJ...681.1646R, 2016NatAs...1E...4A}, and phase offset variations on the order of $\sim0.15$ have been observed \citep{2016NatAs...1E...4A}. Moreover, the coherence timescale in the observations of weather on the transiting hot Jupiter, HAT-P-7b is tens to hundreds of days \citep{2016NatAs...1E...4A}.

To model the albedo and phase offset variability, we use random draws from a Gaussian Process (GP). GPs are used as a non-parametric method of modeling a function in some continuous input space \citep{BDA}. Every realization of a GP is a random variable drawn from a multivariate normal distribution with a mean vector and covariance matrix, where the matrix is constructed from a covariance function that dictates the shrinkage towards the mean and the correlation between pairs of data points. A widely used covariance function is the squared exponential, given by 
\begin{equation}
k(t_i, t_j) = h^2 \exp\left(-\frac{(t_i-t_j)^2}{2 {\tau}^2} \right).
\end{equation}
In the case of modeling an albedo or phase offset variation, $t$ is time, $h^2$ is the maximum covariance of the variation, and $\tau$ is a timescale dictating the smoothness of the variability. $\mathbf{K}$ is the covariance matrix built from the covariance function, given by
\begin{equation}
\mathbf{K} = \begin{bmatrix} 
    k(t_1, t_1) & k(t_1, t_2) & \dots & k(t_1, t_n) \\ 
    k(t_2, t_1) & k(t_2, t_2) & \dots & k(t_2, t_n) \\ 
    \vdots & \vdots & \ddots & \vdots \\ 
    k(t_n, t_1) & k(t_n, t_2) & \dots  & k(t_n, t_n)
\end{bmatrix}.
\end{equation} 

Let $\mathbf{x_A} = (x_{A1}, x_{A2}, ... x_{An})$ be a vector representing the fractional deviation of the albedo from its mean at $n$ discrete timesteps. In other words, 
\begin{equation}
x_{Ai} = x_A(t_i) = \frac{A_g(t_i) - \left<A_g\right>}{\left<A_g\right>}.
\end{equation}
Let $\mathbf{x_{\theta}}$ be the analogous vector for the phase offset. 
We model $\mathbf{x_A}$ and $\mathbf{x_{\theta}}$ as independent multivariate Gaussian random variables:
$\mathbf{x_A} \sim \mathcal{N}(\mathbf{0}, \mathbf{K_A})$ and $\mathbf{x_{\theta}} \sim \mathcal{N}(\mathbf{0}, \mathbf{K_{\theta}})$. The mean vector is zero because $E[x_{Ai}] = E[x_{\theta i}] = 0$.

\begin{figure}
\epsscale{1.25}
\plotone{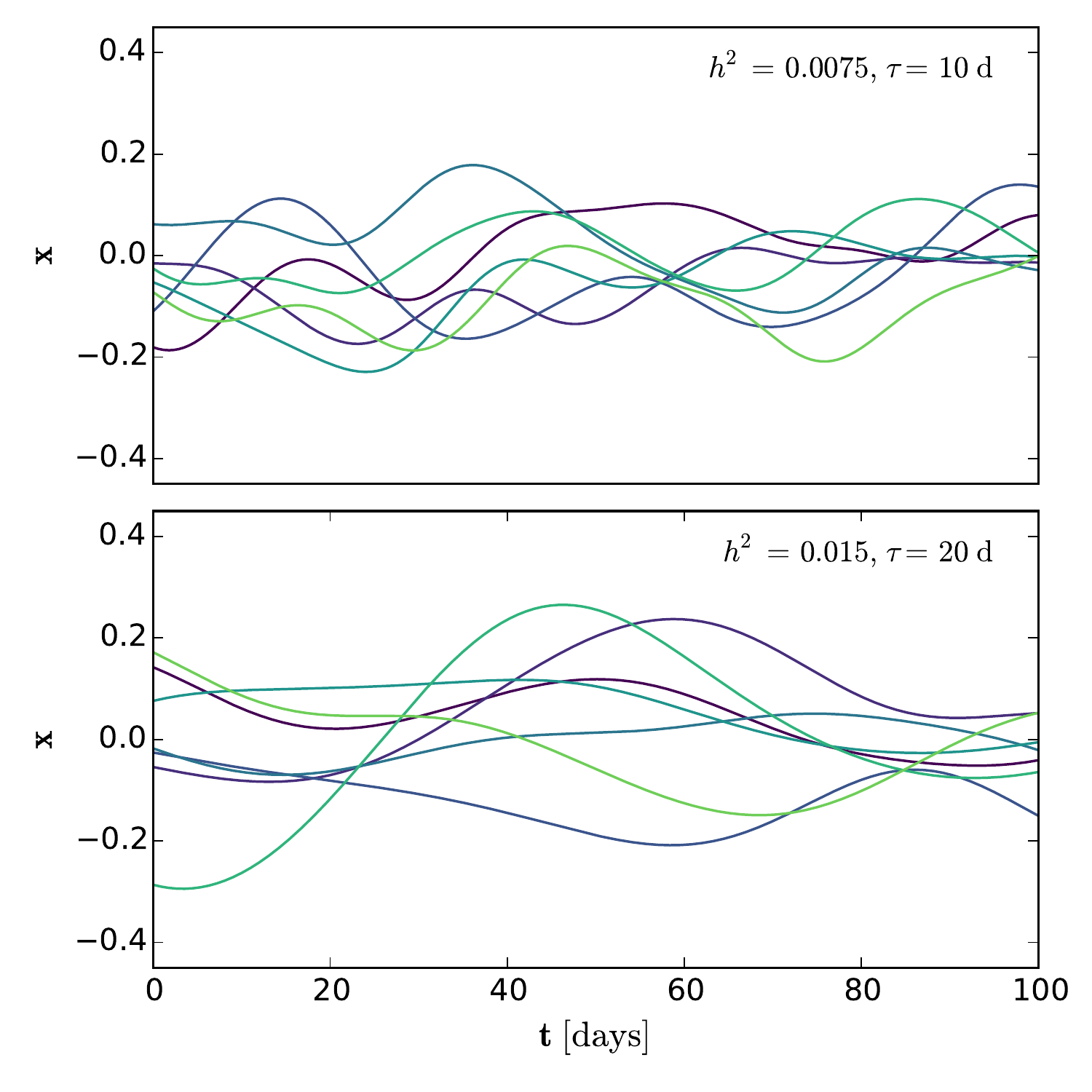}
\caption{Random draws from the GP with different values for the covariance amplitude, $h^2$, and coherence timescale, $\tau$. 
}
\label{albedo variation}
\end{figure}

Figure~\ref{albedo variation} illustrates several realizations of the GP for $h^2 = 0.0075, \tau = 10$ days (top-panel) and $h^2 = 0.015, \tau = 20$ days (bottom-panel). In this figure, $\mathbf{x}$ could represent either $\mathbf{x_A}$ or $\mathbf{x_{\theta}}$.

We construct a synthetic phase curve for each light curve as follows. We draw a set of parameters from the distributions in Table 1. The covariance matrices, $\mathbf{K_A}$ and $\mathbf{K_{\theta}}$, are constructed using the same timescale, $\tau$, but they use their respective covariance amplitudes, ${h_A}^2$ and ${h_{\theta}}^2$. We draw $x_A(t)$ and $x_{\theta}(t)$ as GP realizations. For convenience, we make use of the \textit{george}\footnote{\href{http://dan.iel.fm/george/current/}{http://dan.iel.fm/george/current/}} code for Gaussian Process regression  \citep{hodlr}. Given the random draws for $\left<A_g\right>$ and $\left<\theta\right>$, the albedo variation function is calculated as $A_g(t) = \left<A_g\right>[1 + x_A(t)]$ and the phase offset $\theta(t) = \left<\theta\right>[1 + x_{\theta}(t)]$. These time-dependent quantities and the rest of the parameters are then inserted into the phase curve model outlined in Section~\ref{phase curve model}. Finally, the model phase curve is multiplied by the \textit{Kepler} PDC flux to create the synthetic light curve.

\subsection{Recovery efficiency}

\begin{figure}
\epsscale{1.25}
\plotone{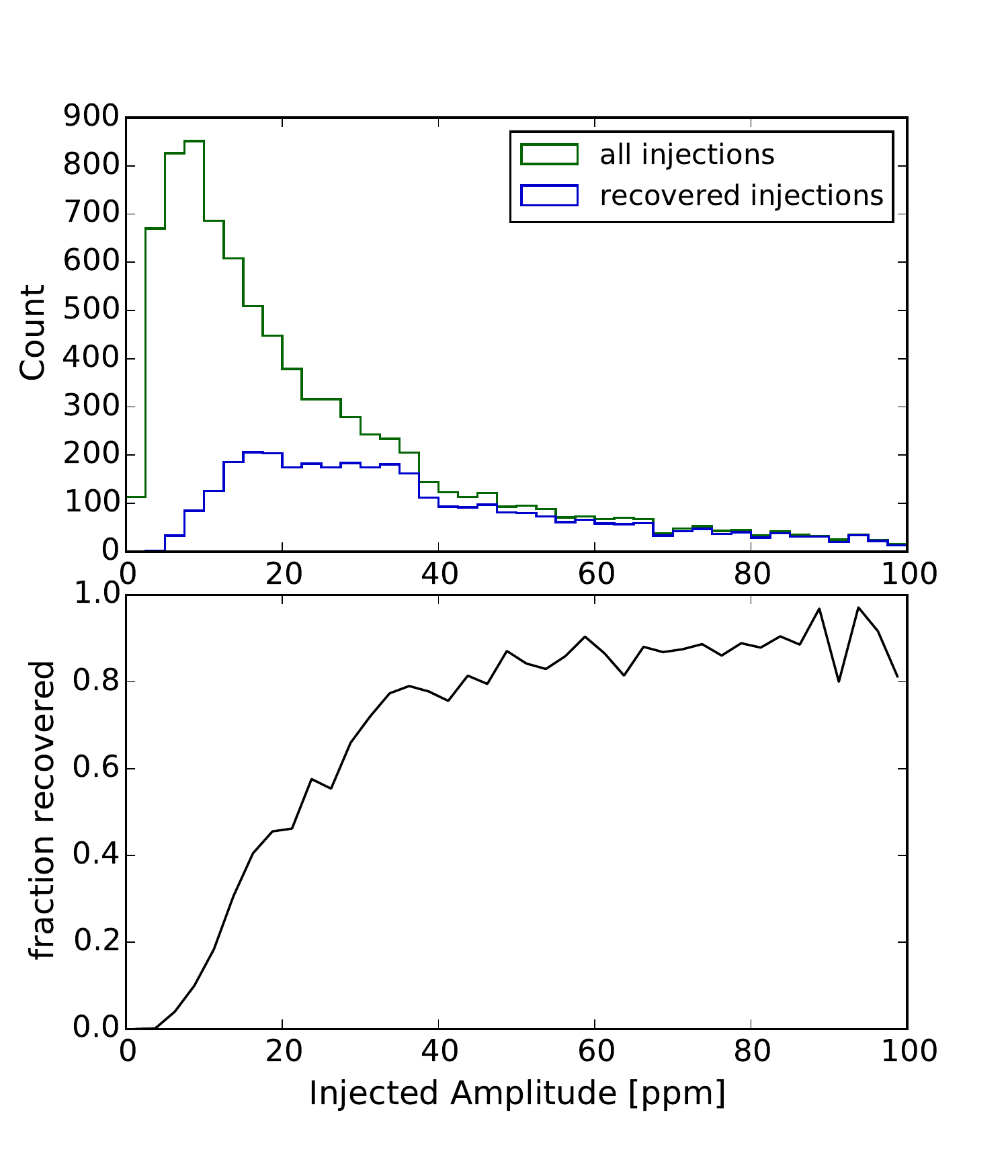}
\caption{The recovery efficiency of the pipeline. In the top panel are histograms of the injected phase curve semi-amplitudes for all injections (in green) and for the injections that were recovered (in blue). In the lower panel is the ratio of the histograms, yielding the fraction recovered as a function of amplitude.}
\label{recovery histograms}
\end{figure}

We now wish to evaluate the efficiency with which these synthetic phase curves are recovered by the pipeline outlined in Section~\ref{Pipeline}. We ran the pipeline on each of the 10,000 synthetic injections. We define the synthetic phase curve to be ``recovered'' when the period of the detected signal is within 0.01 days of the injected signal. In the upper panel of Figure~\ref{recovery histograms}, we plot the histograms of the phase curve semi-amplitudes for all injections and for recovered injections only. The shape of the histogram for the injected data has little physical significance other than to reflect the typical amplitude of the injected phase curves. We show the fraction of recovered injections in the lower panel of the figure. The recovery efficiency is a smooth and increasing function of amplitude, with 20 ppm amplitude phase curves being recovered $\sim 50\%$ of the time.

\begin{figure}
\epsscale{1.25}
\plotone{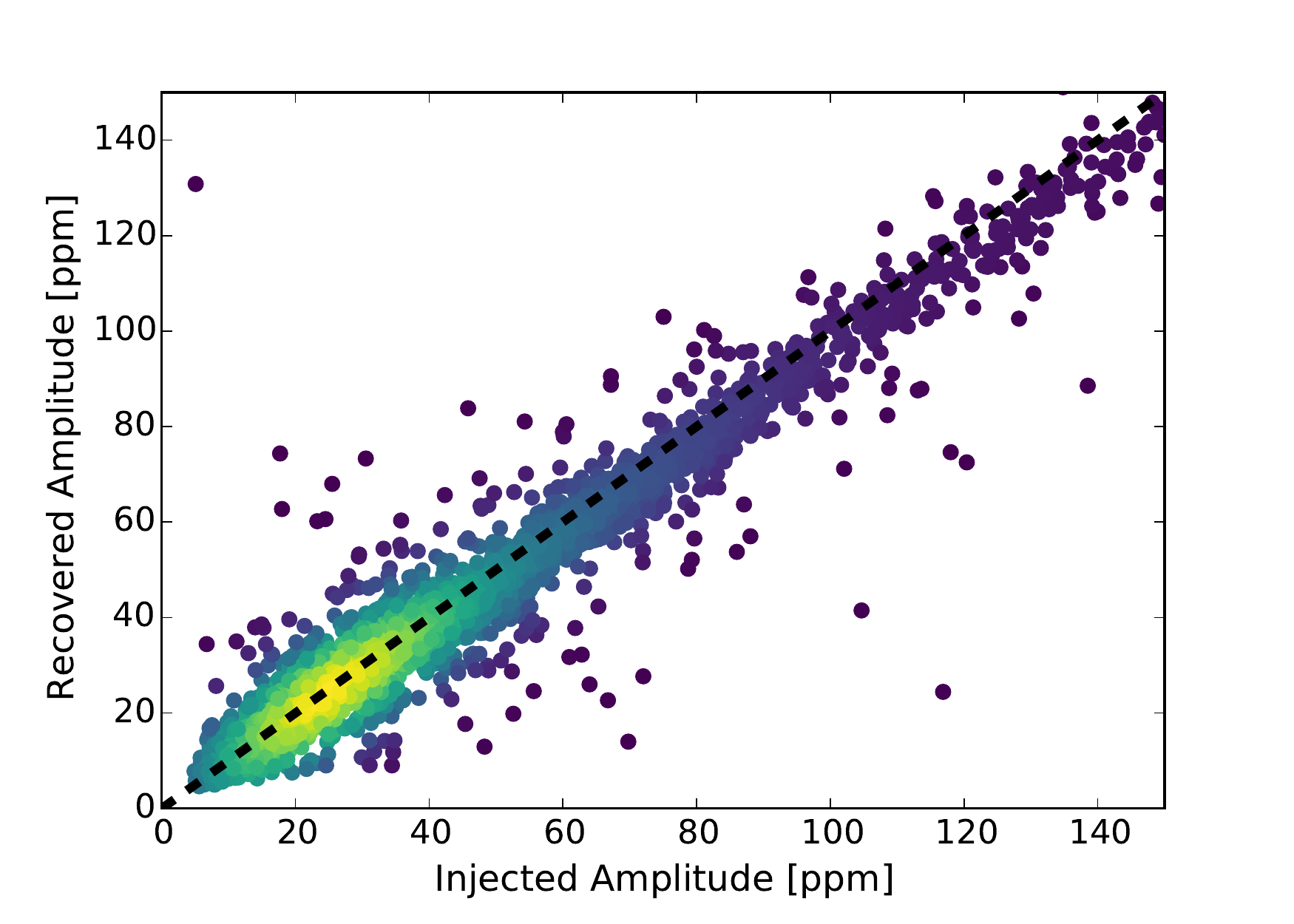}
\caption{For the recovered synthetic phase curve injections, the recovered versus injected semi-amplitudes. The coloration is according to the density of points. }
\label{amplitude recovery}
\end{figure}

For the injections that were recovered, we plot in Figure~\ref{amplitude recovery} the semi-amplitude of the detected signal versus that of the injected signal. The relationship makes clear that the pipeline attains a reasonably accurate recovery of the phase curve amplitudes. The bias towards underestimation at large amplitude is due to the fact that the pipeline uses sinusoidal fits that average over ellipsoidal features in the light curve.

\section{Binary Classification Using Logistic Regression}\label{Logistic Regression}

Having demonstrated that the pipeline is capable of retrieving the signals of synthetic injected phase curves, we proceed to classify light curves by employing a logistic regression algorithm in a supervised machine learning context. We aim to have the algorithm ``learn'' the properties of planetary phase curves and use this information to determine whether phase curves are present in newly introduced light curves. We offer an introduction to logistic regression in Section~\ref{overview} and apply it to phase curve classification in Section~\ref{classification}.

\subsection{Logistic regression overview}
\label{overview}

Logistic regression is a technique used for prediction of a categorical dependent variable based on observed characteristics called explanatory variables or predictors \citep{BDA}. The dependent variable is typically a binary response (e.g. pass/fail, healthy/sick). Like other forms of regression, there may be any number of predictors, and they can be either continuous or discrete.

Consider a dataset of length $n$ containing binary response dependent variables, $Y_i$, which may take on the values 0 or 1. Assume each response variable is associated with $m$ explanatory variables: $x_{1i}, x_{2i}, ... , x_{mi}$. The subscript $i$ indicates an index corresponding to each outcome in the dataset. Let the (unknown) probability of success of the outcome be $p_i$, where $p_i$ is unique to each outcome, but is related to the explanatory variables:
\begin{equation}
E[Y_i \ | \ x_{1i}, x_{2i}, ... , x_{mi}] = p_i.
\end{equation}
The outcomes are thus Bernoulli random variables, 
\begin{equation}
Y_i \ | \ x_{1i}, x_{2i}, ... , x_{mi} \sim \mathrm{Bernoulli}(p_i), 
\end{equation}
with a probability mass function, 
\begin{equation}
\label{p(y_i)}
P(Y_i = y_i \ | \ x_{1i}, x_{2i}, ... , x_{mi}) = {p_i}^{y_i} (1 - p_i)^{1-y_i} .
\end{equation}
In a logistic regression model, we assume that the probability of success varies systematically as a function of the explanatory variables:
\begin{equation}
\label{logit(p_i)}
\mathrm{logit}(p_i) 
= \ln\left(\frac{p_i}{1-p_i}\right) = \pmb{\beta} \cdot \mathbf{x_i}.
\end{equation}
Here, $\pmb{\beta}$ is the vector of regression coefficients,
\begin{equation}
\pmb{\beta} = (\beta_{0}, \beta_{1}, \beta_{2}, ... , \beta_{m}),
\end{equation}
and $\mathbf{x_i}$ is the vector of explanatory variables,
\begin{equation}
\mathbf{x_i} = (1, x_{1i}, x_{2i}, ... , x_{mi}).
\end{equation}
A value of 1 has been appended as the first element corresponding to the intercept coefficient, $\beta_0$. Note that when a given regression coefficient, $\beta_j$, is zero, the outcome is independent of the corresponding explanatory variable, $x_{ji}$.

Rewriting equation~\ref{logit(p_i)} in terms of $p_i$ yields
\begin{equation}
\label{p_i}
p_i = \mathrm{logit}^{-1}(\pmb{\beta} \cdot \mathbf{x_i}) = \frac{1}{1 + e^{-\pmb{\beta} \cdot \mathbf{x_i}}}.
\end{equation}
Using equations~\ref{p(y_i)} and~\ref{p_i}, one can show that the log-likelihood $l(\pmb{\beta})$ can be expressed as
\begin{equation}
\begin{split}
l(\pmb{\beta}) &= \ln \left[ \prod_{i=1}^n p(y_i \ | \ \mathbf{x_i}) \right] \\
&= \sum\limits_{i=1}^n \left[ (1 - y_i) \ln(1 + e^{\pmb{\beta} \cdot \mathbf{x_i}}) -y_i \ln(1 + e^{-\pmb{\beta} \cdot \mathbf{x_i}}) \right]\,.
\end{split}
\end{equation}
The regression is solved via maximum likelihood estimation (MLE) by identifying the vector of regression coefficients, $\pmb{\hat{\beta}}$, that maximize $l(\pmb{\beta})$. In a Bayesian context, one can impose a prior distribution on the regression coefficients, $\pi(\pmb{\beta})$. 
Once the vector of coefficient estimators is found, the unknown probabilities, $p_i$, may be calculated with equation~\ref{p_i}. More importantly, one may estimate the probabilities of \textit{unknown} outcomes if provided the observations of the relevant explanatory variables. 

In the analysis that follows, we perform logistic regression using the 
Scikit-learn library\footnote{\href{http://scikit-learn.org/}{http://scikit-learn.org/}} \citep{scikit-learn}. We use a regularization strength, $\alpha = 1$, and we use Newton-CG optimization for the determination of the regression coefficients. 

\subsection{Phase curve classification using logistic regression}
\label{classification}

We now apply the logistic regression model to the problem of classifying phase curve signals in \textit{Kepler} light curves. The response variable, $Y_i$, is whether the detected signal is a phase curve ($Y_i = 1$) or not ($Y_i = 0$). The probability, $p_i$, is the probability that a given detection is a true planetary phase curve. (Note this is different from the probability that a phase curve is present in the light curve.) 

We used a collection of $m = 12$ predictors, many of them introduced in Section~\ref{Pipeline}:
\begin{enumerate}
\setlength\itemsep{0em}
\item Signal period, $P$
\item Signal amplitude
\item Chi-square ratio of the two-component to one-component phase curve fits, ${\chi^2}_2/{\chi^2}_1$
\item Durbin-Watson statistic ratio, $d_2/d_1$
\item $Q_h$ vs. $h$ slope ratio, $m_2/m_1$
\item $Q_h$ vs. $h$ y-intercept ratio, $b_2/b_1$
\item Full-width half maximum (FWHM) of the peak at $P$ in the Lomb-Scargle periodogram
\item Significance of the peak in the Lomb-Scargle periodogram
\item Significance of the peak in the ``local significance'' periodogram
\item Normalized significance of the $P/2$ harmonic peak in the Lomb-Scargle periodogram
\item Normalized significance of additional peaks in the periodogram
\item Deviation of the phase-folded light curve from sinusoidality
\end{enumerate}
Predictors 1-9 were discussed in Section~\ref{Pipeline}. Here we introduce predictors 10-12, which were all motivated by inspection of the signals in the synthetic datasets. 
\begin{itemize}
    \item[-] Predictor 10: True phase curves should exhibit some power in the periodogram at the $P/2$ harmonic due to the ellipsoidal variation component of the phase curve. We quantify this with predictor 10 by finding the significance of the $P/2$ harmonic peak in the Lomb-Scargle periodogram and taking its ratio with respect to the peak at $P$.
    
    \item[-] Predictor 11: For true phase curve detections, there should not be significant peaks remaining in the periodogram once the signals at $P$ and $P/2$ are removed. For non-phase curve signals such as astroseismic pulsations \citep[e.g.][]{2010ApJ...713L.192G}, significant peaks are accompanied by additional peaks. To calculate predictor 11, we start with the Lomb-Scargle periodogram and remove the peaks at the signal period, $P$, and at the harmonic, $P/2$. We then calculate the local significance periodogram, find the highest peak, and normalize it with respect to the height of the peak at $P$. 
    
    \item[-] Predictor 12: Large-amplitude phase curves should tend to exhibit a deviation from sinusoidality due to the presence of the beaming and ellipsoidal components. We quantify this tendency with predictor 12. We first use the fit in Section~\ref{phase curve fitting} to detrend the light curve. We phase-fold it at the candidate period, bin it, fit a sinusoid to the phase-folded signal, and calculate the residuals of the fit. The predictor is then the Durbin-Watson statistic of the residuals (see Section~\ref{residual autocorrelations}). This quantifies the autocorrelation in the residuals, or the deviation from sinusoidality.

\end{itemize}

\begin{figure}
\epsscale{1.2}
\plotone{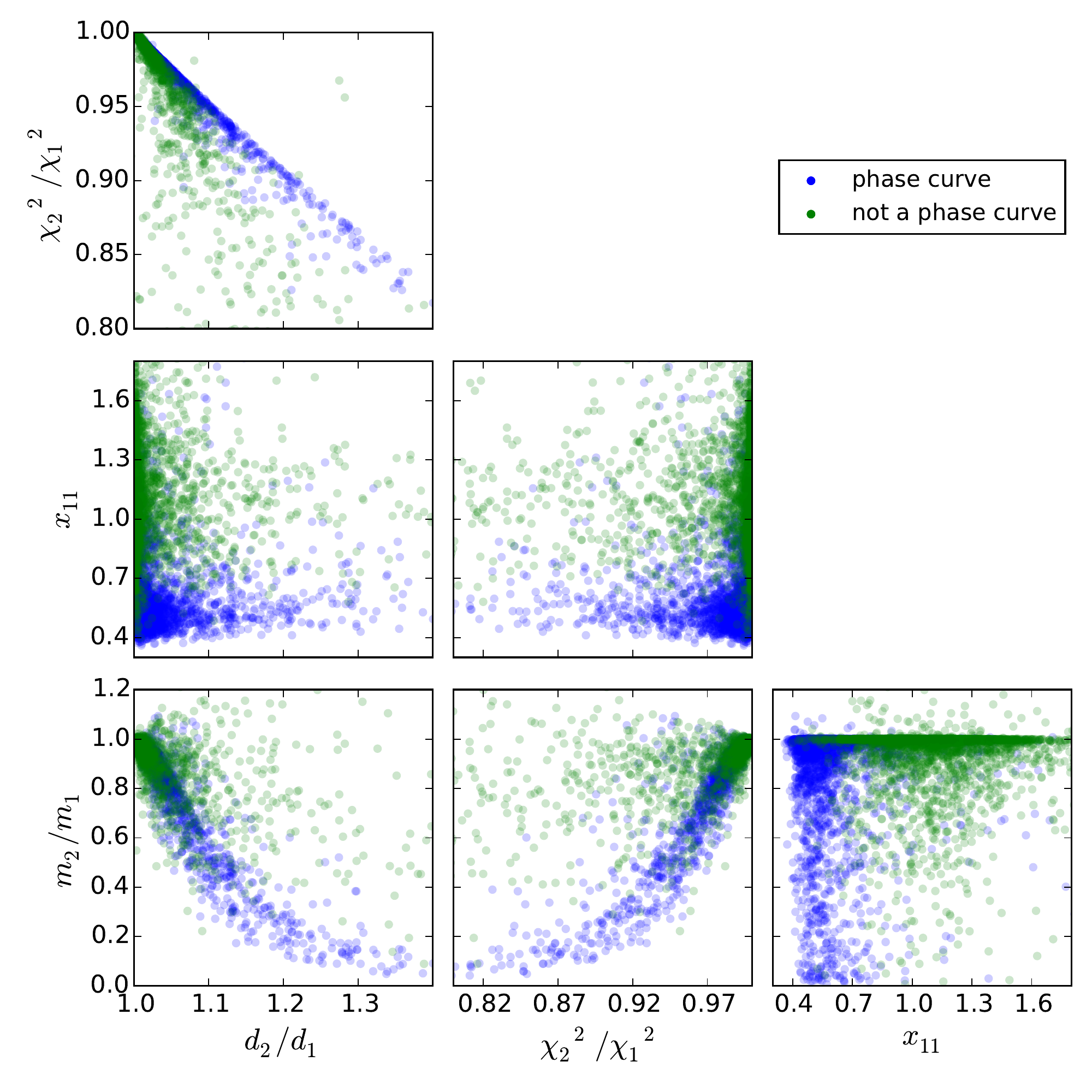}
\caption{A triangle plot showing the correlations between four of the 12 predictors: predictor 3, ${\chi^2}_2/{\chi^2}_1$; predictor 4, $d_2/d_1$; predictor 5, $m_2/m_1$; and predictor 11, $x_{11}$. The points represent the synthetic light curves generated in Section~\ref{light curve injection}. Blue points are recovered phase curve injections. Green points are ``non-phase curves'', that is, detections that did not match the injected phase curve. Broadly speaking, the predictors show very significant differences between the two populations.}
\label{triangle plot}
\end{figure}

All 12 predictors were shown to exhibit significant differences between phase curve detections and non-phase curve detections in the synthetic datasets, either independently or in combination with other predictors. We show an example of this using a triangle diagram in Figure~\ref{triangle plot}. This figure shows the correlations between four of the predictors: predictor 3, ${\chi^2}_2/{\chi^2}_1$; predictor 4, $d_2/d_1$; predictor 5, $m_2/m_1$; and predictor 11, $x_{11}$. Plotted in blue are the recovered phase curve injections in the synthetic data. In green are the detections that did not correspond to the injected phase curve. 

It is clear to see that there are significant differences in the two populations. Consider, for example, the top left plot of ${\chi^2}_2/{\chi^2}_1$ vs. $d_2/d_1$. Both phase curves and non-phase curves exhibit ${\chi^2}_2/{\chi^2}_1 < 1$ because the two-component fit from section~\ref{residual autocorrelations} has more degrees of freedom than the one-component fit. However $d_2/d_1$ for phase curves tends to be larger and follow a straight line because the two-component fit does the best job at removing the residual autocorrelations when the signal is a true phase curve, or a near-sinusoid. (Recall that as the amount of positive autocorrelation decreases,  $d$ increases and $m$ decreases. Therefore, for strong phase curve detections, we should expect $d_2/d_1 \gg 1$ and $m_2/m_1 \ll 1$.) Generally speaking, the separation between the phase curve and non-phase curve populations continues at smaller scales (where the points are overlapping), but it is less pronounced.

Given the distinct differences in the predictors between signals that are or are not phase curves, the multivariate logistic regression should be capable of picking up on these differences and classifying light curves accordingly. We describe this process next.

\subsection{Testing on synthetic phase curves}
\label{synthetic data logistic regression}

We applied the logistic regression to the set of 10,000 synthetic phase curves that we produced in Section~\ref{light curve injection}. We first randomly split the synthetic dataset into a training set and a test set, with the test set being 5\% of the total. We then regressed on the training set using the 12 predictors introduced in the previous section and the outcomes ${Y_i}$ indicating whether each detected signal of the training group was the injected phase curve or not. Finally, we ran predictions on the test set, supplying the 12 predictors and calculating the predicted probabilities that each member of the test set was or was not a phase curve. The predictions were then compared to the actual results, that is, the outcomes of whether each signal detected by the pipeline was the injected phase curve or not.

\begin{figure}
\epsscale{1.3}
\plotone{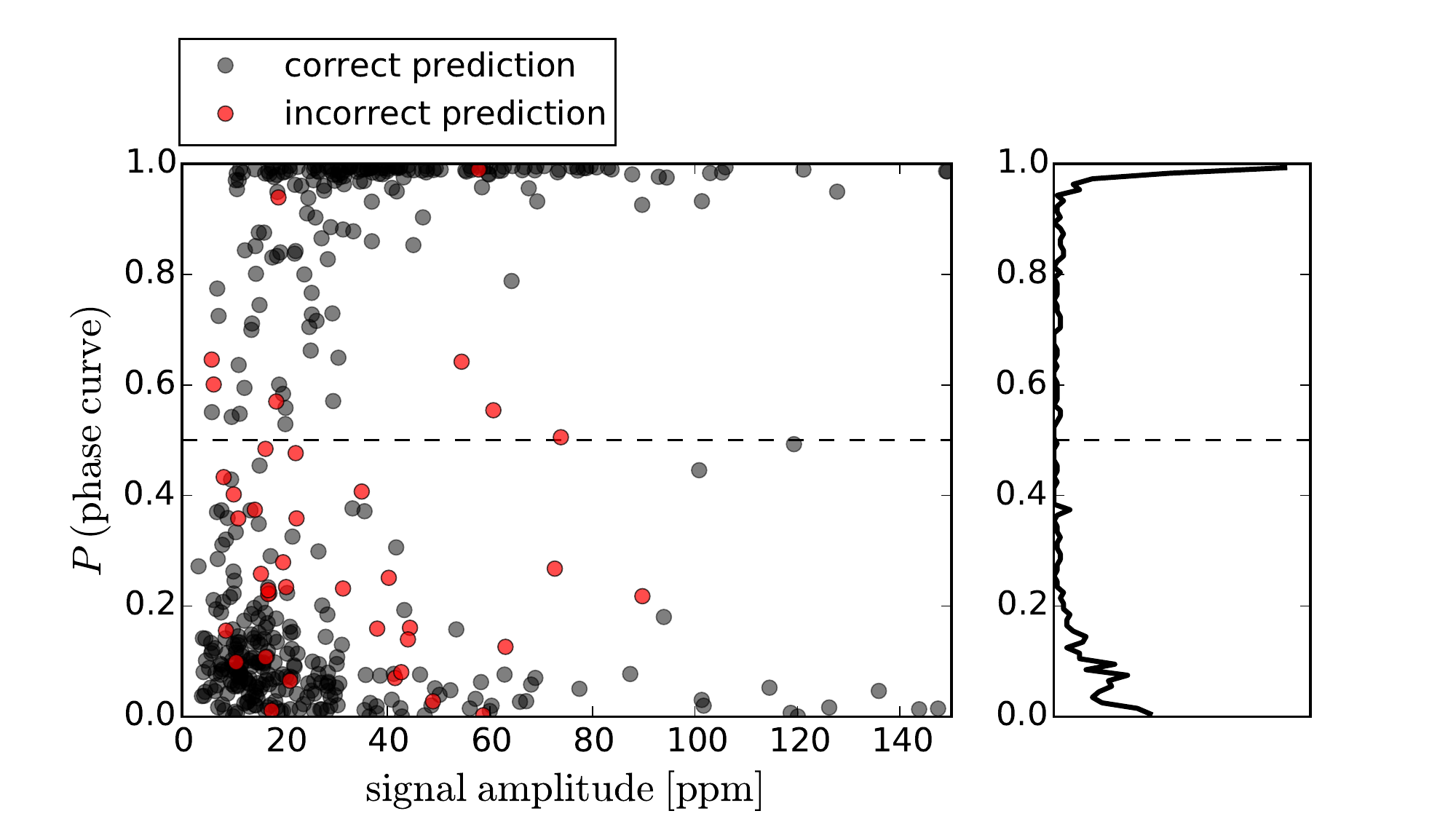}
\caption{The results of the logistic regression on a test subset of the synthetic light curves. On the x-axis is the semi-amplitude of the signal detected by the pipeline in Section~\ref{phase curve fitting}. On the y-axis is the probability predicted by the logistic regression that the detection is a true phase curve. The horizontal dashed line at 0.5 is the boundary between where a signal is predicted to be a phase curve or not. The panel on the right is a histogram of the probabilities of the correct predictions, showing the large number of points with probabilities very close to 0 and 1.}
\label{synthetic data prediction}
\end{figure}

The results of the test are shown in Figure~\ref{synthetic data prediction}. On the x-axis is the semi-amplitude of the signal derived by the two-component fit described in Section~\ref{phase curve fitting}. (Note this amplitude is half the peak-to-peak amplitude.) The y-axis is the probability predicted by the logistic regression that the detection is a true phase curve. Black dots are correct predictions, and red dots are incorrect predictions. More explicitly, black dots with $\mathrm{P(phase \ curve)} > 0.5$ are cases where the logistic regression predicted that the signal detected by the pipeline was a phase curve, and the signal did indeed correspond to the injected phase curve. Red dots with $\mathrm{P(phase \ curve)} > 0.5$ are cases where the logistic regression predicted that the signal was a phase curve, but the signal did not correspond to the injected phase curve. The situation is reversed for points below the $\mathrm{P(phase \ curve)} = 0.5$ line.

Generally speaking, the logistic regression performs remarkably well. In total, correct predictions are made $\sim90\%$ of the time, and the accuracy increases steeply as $\mathrm{P(phase \ curve)}$ approaches 0 or 1. Larger amplitude detections are assigned more definitive probabilities (i.e. $\mathrm{P(phase \ curve)}$ closer to 0 or 1).

\begin{figure}
\epsscale{1.2}
\plotone{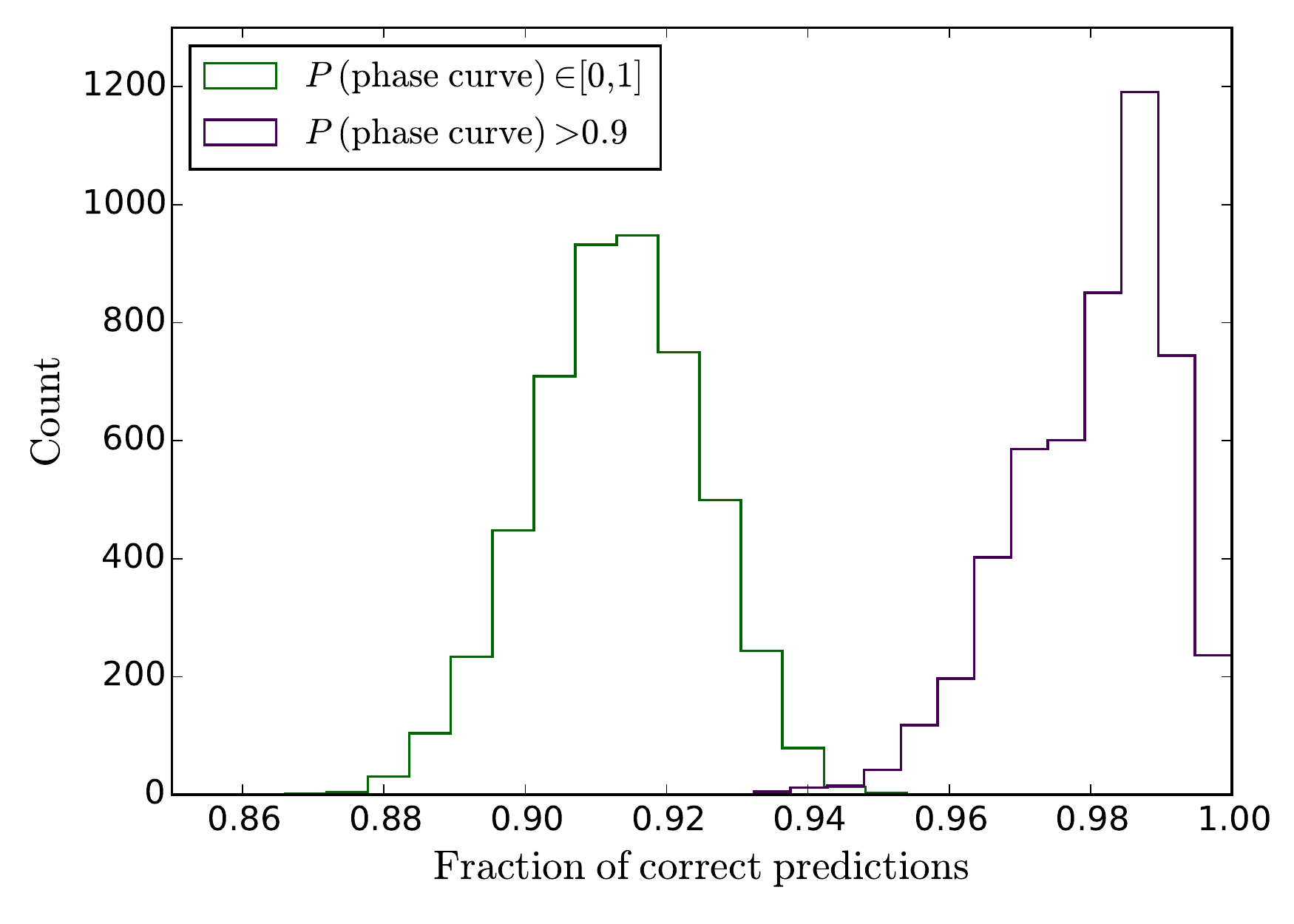}
\caption{Histograms of the fraction of correct predictions for 5,000 random realizations of the synthetic data training and testing procedure. The green histogram shows each realization's fraction of predictions that are correct. The purple histogram is restricted to predictions with probabilities greater than 0.9.}
\label{logistic regression trial histograms}
\end{figure}

We tested the consistency of the algorithm's predictive performance using 5,000 realizations of the training and test procedure described in the first paragraph of this section. Each realization used a random partition of the synthetic dataset into the 95\% training and 5\% test sets. For each trial, we performed the logistic regression and identified the correct and incorrect predictions. Histograms of the fraction of correct predictions are shown in Figure~\ref{logistic regression trial histograms}. The green histogram shows that the algorithm predicts correctly $\sim 91\%$ of the time. For cases with reported probabilities greater than 0.9 (purple histogram), the prediction is correct $\sim 98\%$ of the time. As indicated by the widths of the histograms, these results are fairly consistent from trial to trial.

\begin{figure}
\epsscale{1.3}
\plotone{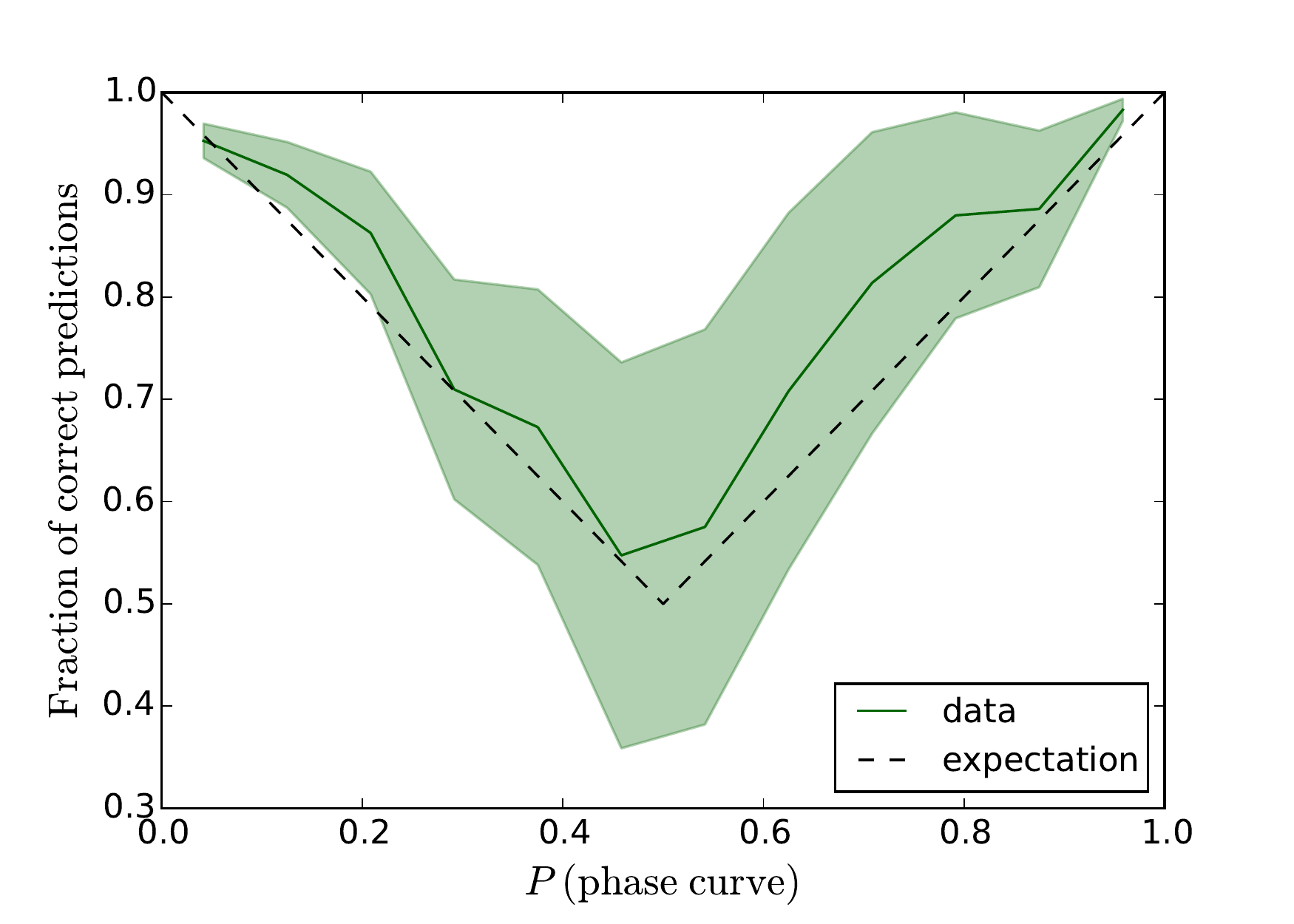}
\caption{The fraction of correct predictions versus the probability reported by the logistic regression. The dark green curve is the mean over the 5,000 realizations, and the surrounding lighter green band shows the standard deviation. The standard deviation is largest towards the center because most points in a given realization are predicted with probabilities near 0 or 1 (see Figure~\ref{synthetic data prediction}). The dashed black line is the probability theory expectation.}
\label{frac correct versus probability}
\end{figure}

We also computed the fraction of correct predictions as a function of the probability reported by the logistic regression. The mean curve over the 5,000 random realizations is displayed in Figure~\ref{frac correct versus probability}. The results indicate that the reported probabilities can indeed be interpreted as the likelihood that a given signal corresponds to a true phase curve.

\subsection{Testing on transiting \textit{Kepler} hot Jupiters}
\label{testing on transiting data}

We also tested the detection pipeline and logistic regression on the collection of transiting hot Jupiters in the \textit{Kepler} field. We first compiled the list of KOIs in the Q1-Q17 DR 24 catalog \citep{2016ApJS..224...12C} from the NASA Exoplanet Archive\footnote{\label{NASA Exoplanet Archive}\href{http://exoplanetarchive.ipac.caltech.edu}{http://exoplanetarchive.ipac.caltech.edu}} \citep{2013PASP..125..989A}. We filtered the confirmed planets and planet candidates to those with periods in a 1-5 day range and radii in the range 0.5-3 $R_{\mathrm{Jup}}$. For the unconfirmed planet candidates, we used the KOI false positive probabilities (FPPs) calculated by \cite{2016ApJ...822...86M} to restrict the list to those with a FPP less than 10\%. These steps resulted in a collection of 59 total hot Jupiter planets and planet candidates with potentially detectable phase curves.

The pipeline described in Section~\ref{Pipeline} was executed in the same manner as on the synthetic datasets. The only difference was that the transits and (potential) secondary eclipses were removed from the light curves before processing them. The pipeline recovered planetary phase curves in four cases: HAT-P-7b (also called Kepler-2b), Kepler-13b, Kepler-41b, and Kepler-76b.  In these cases, the period of the signal recovered by the pipeline matched the planet period to within 0.01 days. This apparently low recovery rate is not surprising. \cite{2015ApJ...804..150E} conducted a comprehensive search for planetary phase variations in \textit{Kepler} transiting planets, finding 14 planets total. Twelve of these have parameters fitting our criteria ($1 \ \mathrm{day} < P < 5 \ \mathrm{days}$, $0.5 \ R_{\mathrm{Jup}} < R_{\mathrm{P}} < 3 \ R_{\mathrm{Jup}}$). In this sense, there are 12 potentially detectable phase curves, and one-third of these have amplitudes large enough to be detected by our pipeline. 

Recovery by the pipeline is distinct from recovery by the logistic regression. The next step is to consider predictions of the transiting dataset to confirm that the logistic regression places high probabilities on the four recovered phase curve signals and low probabilities on the rest. 

\begin{figure}
\epsscale{1.25}
\plotone{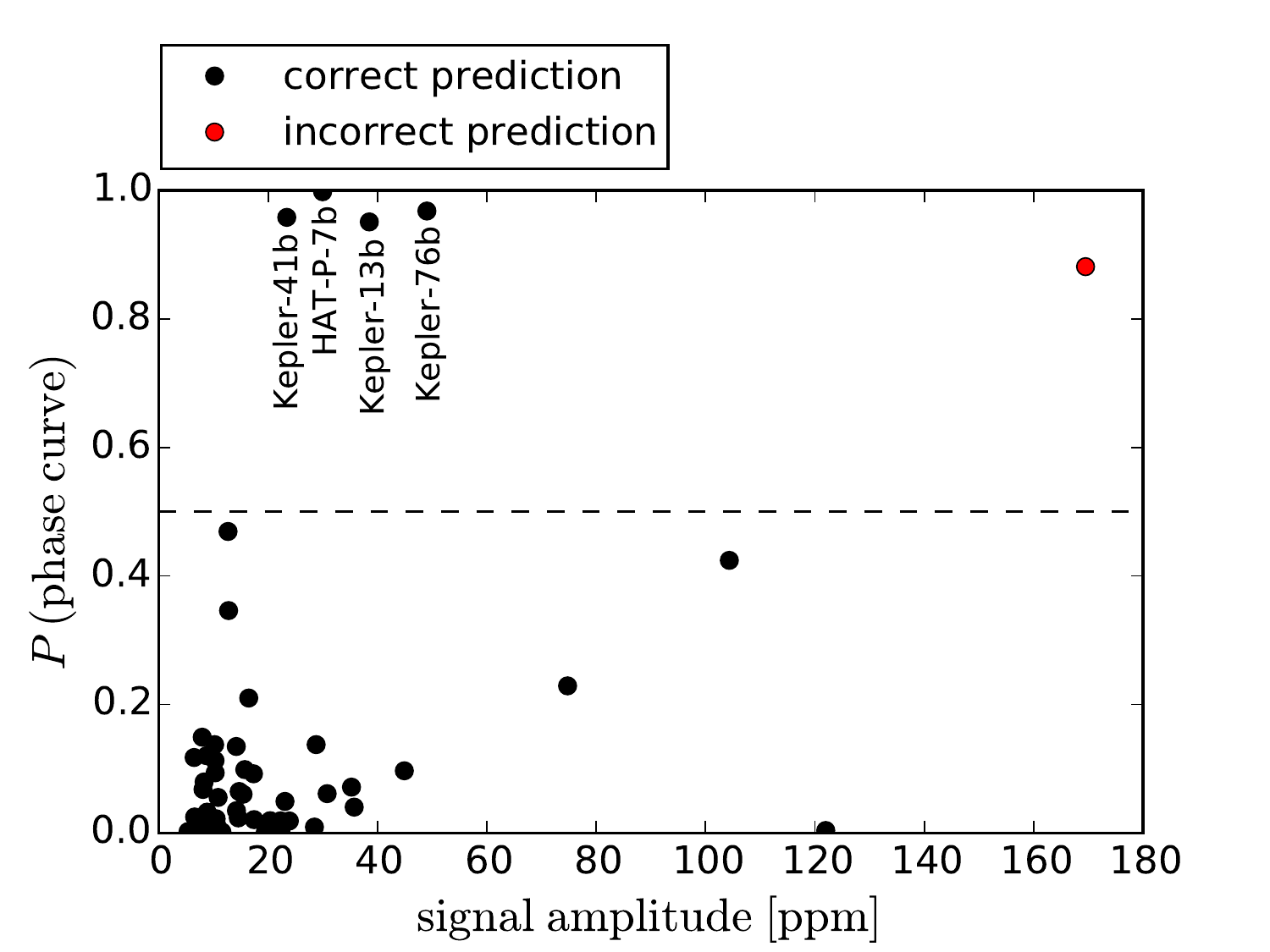}
\caption{The results of the logistic regression on a set of transiting planets in the \textit{Kepler} field with $1 \ \mathrm{day} < P < 5 \ \mathrm{days}$ and $0.5 \ R_{\mathrm{Jup}} < R_{\mathrm{P}} < 3 \ R_{\mathrm{Jup}}$. On the x-axis is the semi-amplitude of the signal detected by the pipeline in Section~\ref{phase curve fitting}. On the y-axis is the probability predicted by the logistic regression that the detection is a true phase curve. The horizontal dashed line at 0.5 is the boundary between where a given signal is predicted to be a phase curve or not. The four annotated black points with $P$(phase curve) $> 0.5$ correspond to the cases where pipeline recovered the planetary phase curve, and the logistic regression correctly identified these signals to be true phase curves. The black points with $P$(phase curve) $< 0.5$ are cases where the signal detected by the pipeline did not correspond to the planetary period, and the logistic regression correctly predicted that the signals were not true phase curves.}
\label{transiting data prediction}
\end{figure}

Using the regression that was performed on the synthetic data in the previous section, we made predictions on the light curves of the transiting planets. The results are shown in Figure~\ref{transiting data prediction}. In all four cases where the pipeline recovered the planet's phase curve, the logistic regression correctly predicted that the detection was a true phase curve. This is represented by the four annotated black points with $\mathrm{P(phase \ curve)} > 0.5$. In all but one of the remaining cases, the logistic regression correctly predicted that the signal detected by the pipeline was not a phase curve. Put another way, all of the black points with $\mathrm{P(phase \ curve)} < 0.5$ were cases where the pipeline did not recover the planet's period, and the logistic regression correctly reported that the signals were not real phase curves.  

The one false positive pictured in the upper right of Figure~\ref{transiting data prediction} has an amplitude greater than that expected from a planetary phase curve. The signal is from the light curve of KIC 5651104/KOI-840/Kepler-695b. UKIRT images available from the \textit{Kepler} Exoplanet Follow-up Observing Program (ExoFOP)\footnote{\href{https://exofop.ipac.caltech.edu/}{https://exofop.ipac.caltech.edu/}} show that the star has nearby background or foreground companions, so the signal probably results from blended, non-eclipsing binary stars. We found further evidence for this interpretation in that the phase curve amplitude is inconsistent when the light curve is folded at two times the detected period. Details on the nature of this inconsistency and steps taken to rule out these types of false positives will be discussed in Section~\ref{binary star filtering}.\\

In short, the logistic regression performs exceedingly well on both synthetically generated datasets and the light curves of transiting $\textit{Kepler}$ planets. In the vast majority of cases, the algorithm correctly predicts when a signal is or is not a true planetary phase curve. This gives us confidence in the process of applying the logistic regression to novel light curves and inspecting them for phase curve detections.

\section{Application to \textit{Kepler} FGK stars}

We now describe the application of our pipeline and logistic regression algorithm to the search for non-transiting planets around \textit{Kepler} FGK stars without confirmed planets or KOIs. We first downloaded the Q1-Q17 DR 25 \textit{Kepler} stellar catalog \citep{2017ApJS..229...30M} from the NASA Exoplanet Archive$^{\ref{NASA Exoplanet Archive}}$ \citep{2013PASP..125..989A}. We filtered the target stars for FGK main sequence stars, using the criteria from \cite{2016ApJ...828...99C}: $4000 \ K < T_{\mathrm{eff}} < 7000 \ K$ and $\mathrm{log}\,g > 4.0$. We then removed stars with known planets and planetary candidates, leaving 146,980 targets. For these targets, we retrieved all 17 quarters of \textit{Kepler} long-cadence, pre-search conditioning (PDC) photometry. These light curves are publicly available at the Mikulski Archive for Space Telescopes (MAST). 

We then ran the pipeline outlined in Section~\ref{Pipeline} on the set of light curves, with successful convergence on 142,630 of them. This procedure identified the most likely phase curve signal in each light curve, if a phase curve were to be present. Using the logistic regression that was performed on the synthetic data in Section~\ref{synthetic data logistic regression}, we then made predictions on this novel set of light curves, evaluating the likelihood that each detected signal could correspond to a true phase curve. 

\begin{figure}
\epsscale{1.25}
\plotone{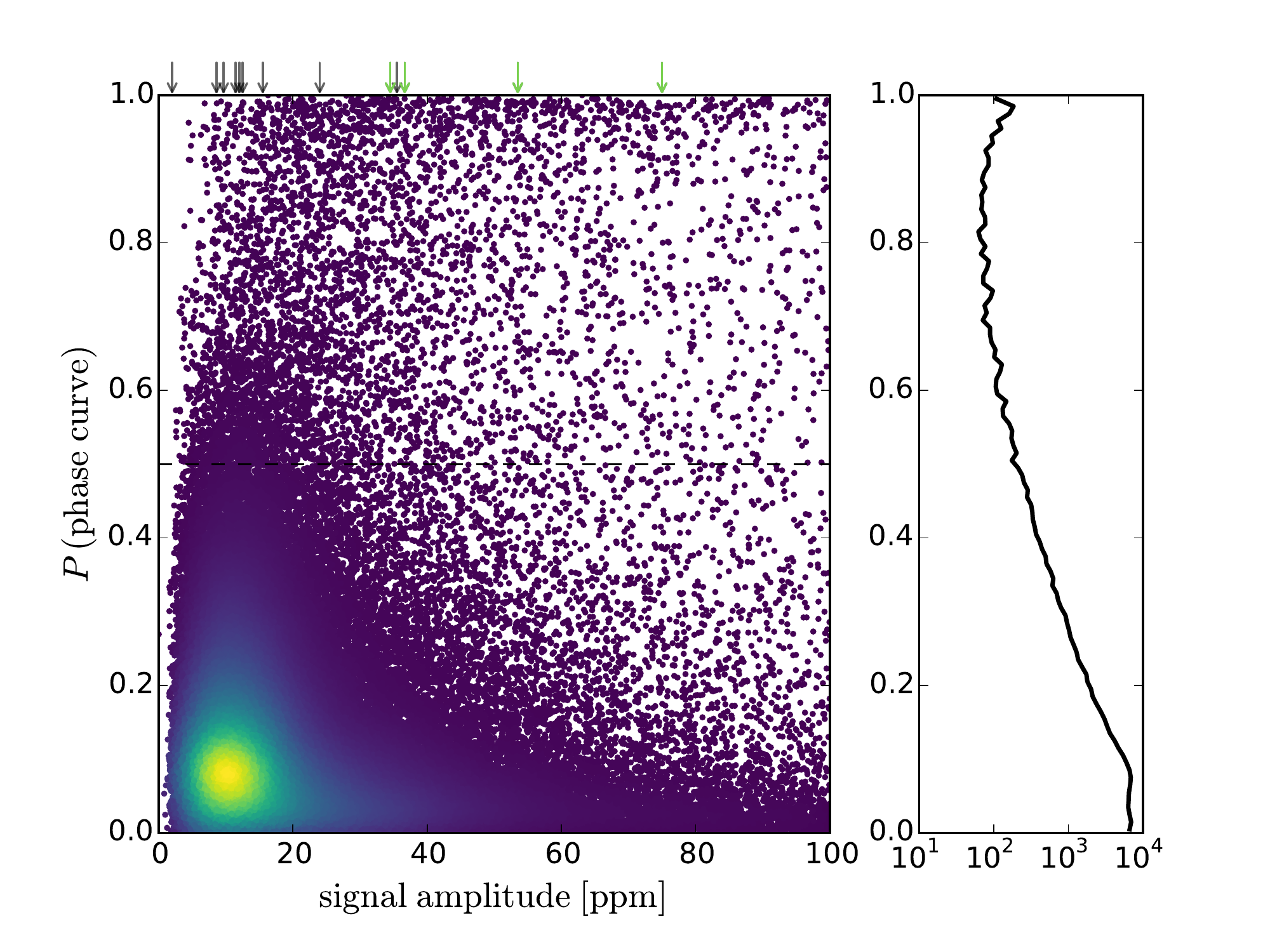}
\caption{The results of the logistic regression on a set of 142,630 \textit{Kepler} FGK main sequence stars without known planets or planet candidates. On the x-axis is the semi-amplitude of the signal detected by the pipeline in Section~\ref{phase curve fitting}. On the y-axis is the probability predicted by the logistic regression that the detection is a true phase curve. The points have been colored according to their density. The downward pointing arrows are located at the phase curve semi-amplitudes of known transiting planets from \cite{2015ApJ...804..150E}, with the green arrows being the four planets recovered in Section~\ref{testing on transiting data} (see Figure~\ref{transiting data prediction}). The panel on the right is a log-scale histogram of the probabilities, showing that the vast majority of points have probabilities less than 0.5. }
\label{real data prediction}
\end{figure}

The results are displayed in Figure~\ref{real data prediction}. This plot is analogous to Figures~\ref{synthetic data prediction} and~\ref{transiting data prediction}, with the x-axis being the semi-amplitude of the signal derived by the two-component fit described in Section~\ref{phase curve fitting} and the y-axis being the probability output by the logistic regression described in Section~\ref{classification}. The points have been colored by their density, and the panel at the right is a histogram of the probabilities on a log scale. As expected and required, the vast majority of targets have $P(\mathrm{phase \ curve}) < 0.5$.

\subsection{Binary star filtering}
\label{binary star filtering}
Among the potential sources for astrophysical false positives, short period binary stars are one of the most significant concerns. Binary stars have an occurrence rate of $\gtrsim 10\%$ at periods of 1-5 days \citep{2016AJ....151...68K}, about two orders of magnitude greater than the occurrence rate of hot Jupiters with phase variations detectable by our pipeline. Their phase curves are dominated by ellipsoidal variations \citep{2012ApJ...746..185F}, meaning that our pipeline would detect them at half their orbital period. Although the amplitudes of their ellipsoidal variations are typically 100-1000 ppm, the amplitude is given by $A_{\mathrm{ellip}} \propto \sin^2i$ (see equation~\ref{ellipsoidal variation}), so inclined systems could appear with amplitudes less than 100 ppm. Moreover, potential background or foreground stars could contaminate the photometric aperture and make the ellipsoidal variations appear with small amplitudes.

Fortunately, there are means of discriminating between a binary system and a giant planet phase curve, even at small amplitudes. Because ellipsoidal variations dominate the binary stars' light curve, they would be detected at half their orbital period. However, the Doppler beaming amplitude is non-negligible \citep{2012ApJ...746..185F}, meaning that the periodogram of a binary light curve should show significant power at the true period, or twice the detected period. Planetary phase curves should not show any signal at $2P$. Furthermore, when the light curve is folded at twice the \textit{detected} period and split at the midpoint, the beaming amplitude of a binary system will cause the two halves to be distinct, but a planetary signal will have two identical curves. To address these features and eliminate binary stars, we constructed the following three filters. The values of the thresholds were determined by inspection of the synthetic datasets.
\begin{enumerate}
\item Require $\mathrm{power}_{2P}/\mathrm{power}_{P} < 0.4$, where $P$ is the detected period (or, in the case of binaries, half the true period).
\item If the light curve is folded and binned at $2P$, and if $y_i$ represents the data in the first half and $z_i$ represents the second half, then require $0.8 < r < 1.2$, where 
$r = \frac{\frac{1}{N}\sum_{i=1}^{N}(y_i - z_i)^2}{{\sigma_y}^2 + {\sigma_z}^2}$, ${\sigma_y}^2 = \sum_{i=1}^{N}{\sigma_{yi}}^2$ and ${\sigma_z}^2 = \sum_{i=1}^{N}{\sigma_{zi}}^2$. We take $N=200$, such that 400 points span the $2P$ folded light curve.
\item Reqiure $\frac{\mathrm{Amp}}{300 \ \mathrm{ppm}} < \left(\frac{P}{1 \ \mathrm{day}}\right)^{-4/3}$, where Amp is the semi-amplitude.
\end{enumerate}

To apply these filters to the sample, we eliminated all points with $P > 0.5$ that did not adhere to these limits. The results of these filters are shown in Figure~\ref{before and after filters}, where they are discussed in conjunction with the additional vetting in the next section.

\subsection{Further vetting via physical plausibility considerations}
\label{candidate vetting}

In addition to binary stars, there are likely additional types of false positives that are not being taken into account \citep{2017arXiv170300496S}. It is therefore worthwhile to take further steps to address these.

Another powerful method for false positive vetting is to consider whether the signal morphologies are physically plausible given the expectations for planetary phase curves. Similar to the binary star filtering in the previous section, the additional vetting that we will now describe was only applied to cases with predicted probabilities, $P > 0.5$ (hereafter called ``candidates'').

\begin{figure}
\epsscale{1.25}
\plotone{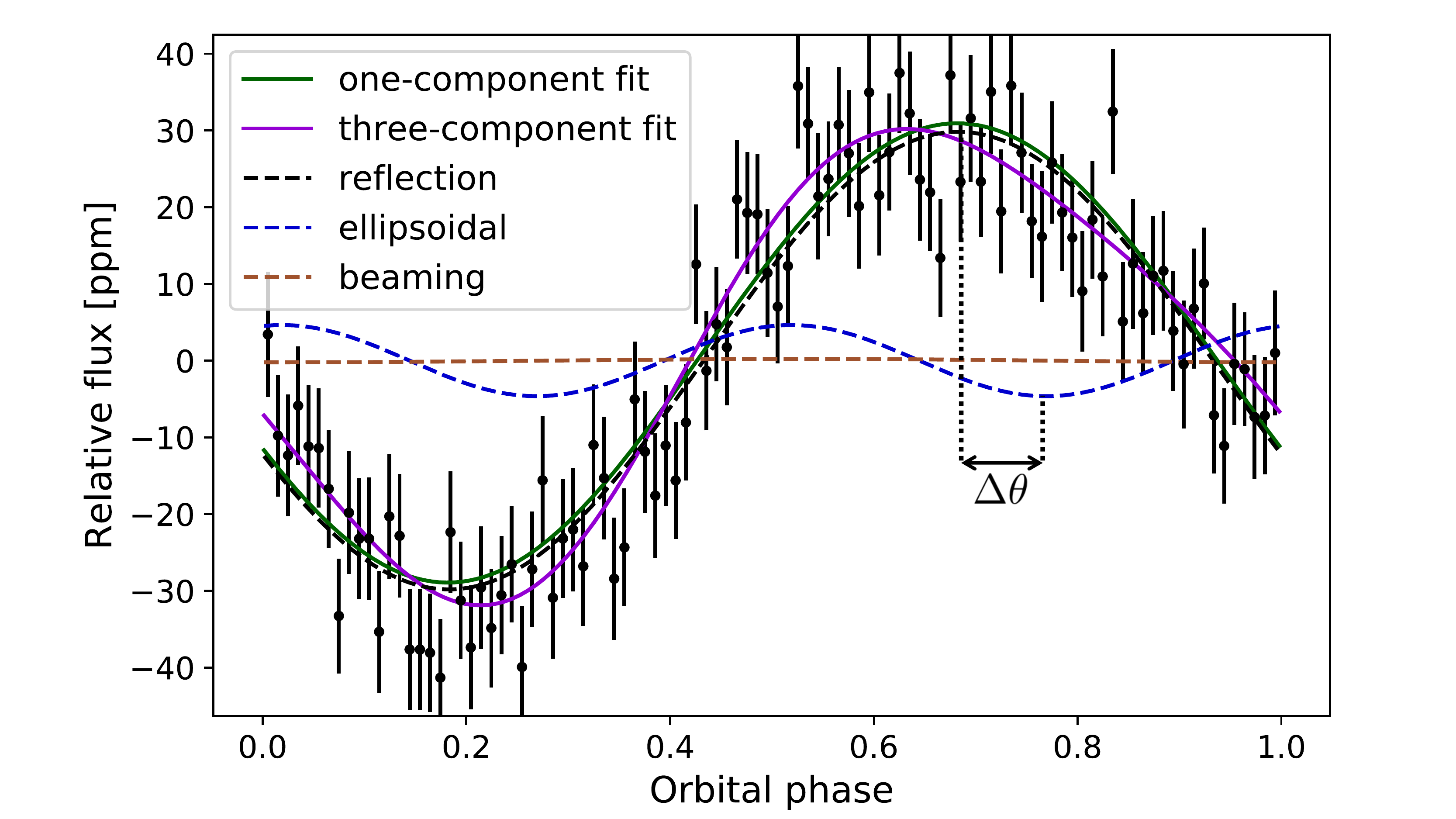}
\caption{An example of the one-component fit (in green) and three-component fit (in purple) for KIC 5716330, one of the candidates. The reflected light, ellipsoidal, and beaming curves composing the three-component fit are displayed with dashed lines, and the phase offset, $\Delta \theta = \theta_{\mathrm{refl}} - \theta_{\mathrm{ellip}}$, is labeled. The fit indicates an eastward shift in the phase curve maximum, with the peak before the substellar point.}
\label{three component fit}
\end{figure}

 We started with a fitting procedure for assessing the physical plausibility of the candidate phase curves. For each candidate, we calculated a one-component sinusoidal fit, representing a fit for reflected light only. We then calculated a three-component fit modeled by 
\begin{equation}\label{three component model}
\begin{split}
f(\phi) = 
1 - A_{\mathrm{refl}} \cos(2\pi(\phi + \theta_{\mathrm{refl}})) \\
- A_{\mathrm{ellip}} \cos(4\pi(\phi + \theta_{\mathrm{ellip}}))  \\ + A_{\mathrm{beam}}\sin(2\pi(\phi + \theta_{\mathrm{ellip}}))\,.
\end{split}
\end{equation}
This is a simple way to assess the phase curve without applying the full phase curve model described in Section~\ref{phase curve model}. The $A_{\mathrm{refl}}$ term represents the reflected light component with one phase, $\theta_{\mathrm{refl}}$, and the $A_{\mathrm{ellip}}$ and $A_{\mathrm{beam}}$ terms are the ellipsoidal and beaming components with a separate phase, $\theta_{\mathrm{ellip}}$. The thermal component is incorporated into the $A_{\mathrm{refl}}$ term, since the two cannot be separated without more advanced modeling. An example of the one-component and three-component phase curve fits for one of the candidates, KIC 5716330, is shown in Figure~\ref{three component fit}. 

We then used these light curve fits to examine the candidates for false positives. We first calculated the Durbin-Watson statistic, $d$, of the residuals of the three-component fits (see equation~\ref{Durbin-Watson}). If the statistic is far from 2, this indicates autocorrelated residuals and implies that the light curve is not well-described by the fit. We eliminated all candidates with $d < 1.85$ or $d > 2.15$, where this range was determined by inspection of the synthetic phase curves.

Next, we examined the candidate phase curves' physical plausibility by considering $A_{\mathrm{refl}}$ and $A_{\mathrm{ellip}}$. If, for example, $A_{\mathrm{refl}}$ is too large, the planet would have an abnormally large radius or albedo. If $A_{\mathrm{ellip}}/A_{\mathrm{refl}}$ is too large, the density would be abnormally high. 

To convert $A_{\mathrm{refl}}$ and $A_{\mathrm{ellip}}$ into physical quantities, we first solve for the planet's radius in terms of $A_{\mathrm{refl}}$ and other observables:
\begin{equation}
R_p =  \left(\frac{2 A_{\mathrm{refl}}}{A_g}\right)^{1/2} \left(\frac{G M_{\star}}{4\pi^2}\right)^{1/3} P^{2/3} (\sin i)^{-1/2}.
\end{equation}
The factor of 2 in front of $A_{\mathrm{refl}}$ comes from the fact that $A_{\mathrm{refl}}$ is a semi-amplitude but equation~\ref{Ap equation} considers a peak-to-peak amplitude.

Next, we solve for the planet's mean density in terms of $A_{\mathrm{ellip}}/{A_{\mathrm{refl}}}^{3/2}$:
\begin{equation}
\rho_p = \frac{\rho_{\star}}{2\sqrt{2}} \left(\frac{A_{\mathrm{ellip}}}{{A_{\mathrm{refl}}}^{3/2}}\right) \left(\frac{{A_g}^{3/2}}{\alpha_2}\right) \left(\sin i\right)^{-1/2}.
\end{equation}
Here $\alpha_2$ is a constant in the ellipsoidal variation model (see equation~\ref{ellipsoidal variation}). It varies nearly linearly with $T_{\mathrm{eff}}$ (see Figure~\ref{alpha constants}).

\begin{figure}
\epsscale{1.3}
\plotone{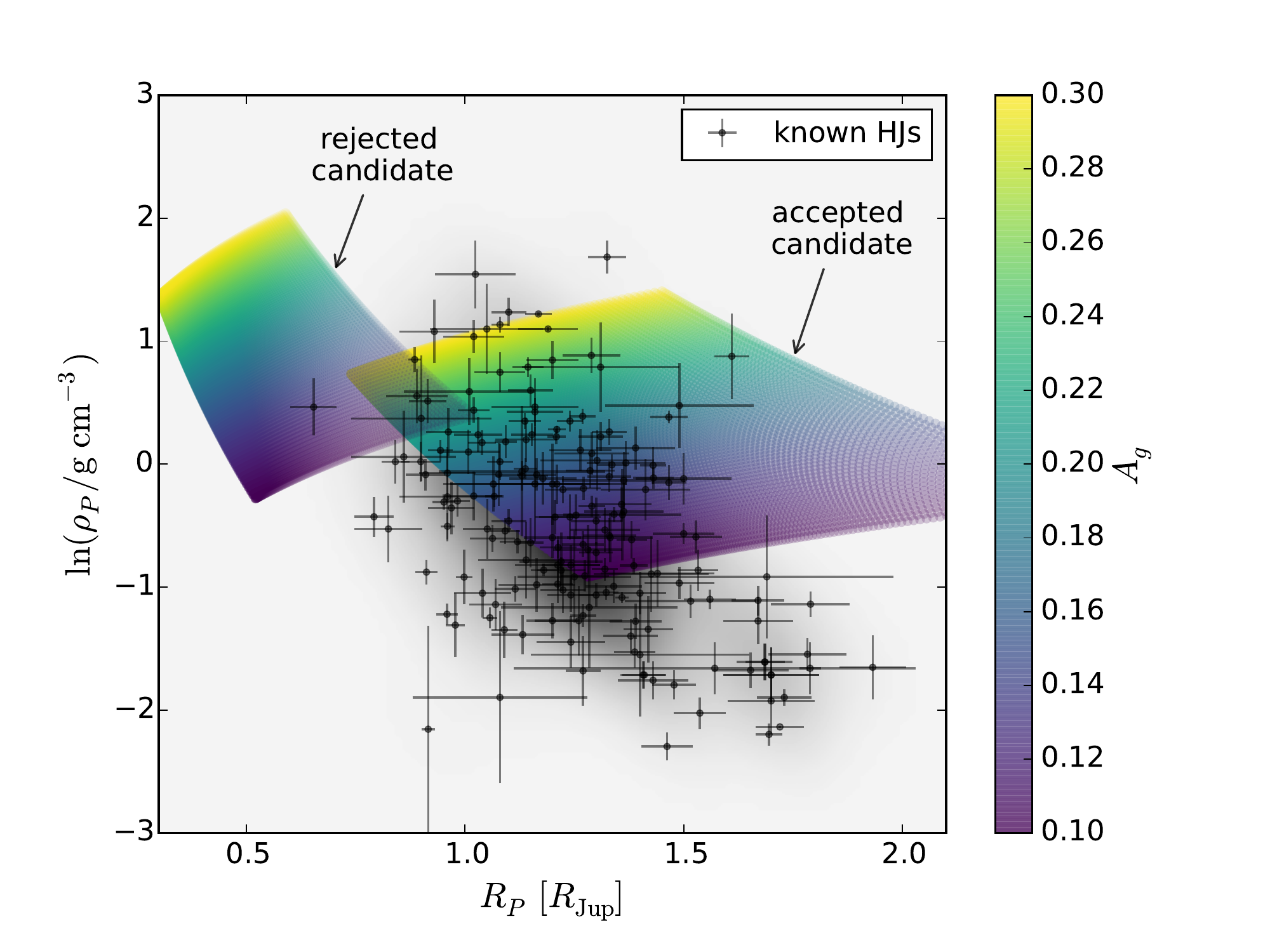}
\caption{The $\rho_p-R_p$ diagram that is used for additional candidate vetting. The black points with errorbars correspond to known hot Jupiters with measured masses and radii; a kernel density estimation corresponding to these data points is shown with the gray gradient. The $\rho_p-R_p$ regions of two candidates are displayed, where $A_g \in [0.1-0.3]$ and $i \in [15^{\circ}, 85^{\circ}]$. The coloration is according to the albedo. The candidate labeled ``accepted'' is physically plausible since its region agrees with known hot Jupiter radii and densities. (It has a large value of $S$.) The candidate labeled ``rejected'' is less plausible.}
\label{rho vs R plot}
\end{figure}

For each candidate, we calculated estimates of $R_p$ and $\rho_p$ by using the \textit{Kepler} stellar catalog properties and by assuming $A_g \in [0.1-0.3]$ and $i \in [15^{\circ}, 85^{\circ}]$. This domain thus corresponds to an allowable region in $\rho_p-R_p$ space. The candidate's plausibility may be considered by comparing this region to the radii and density of known hot Jupiters. In Figure~\ref{rho vs R plot}, we plot $\ln\rho_p$ vs. $R_p$ for all hot Jupiters with measured masses and radii, as obtained by the Exoplanets Data Explorer\footnote{\href{exoplanets.org}{exoplanets.org}} \citep{2014PASP..126..827H}. (We considered planets with $P < 5 \ \mathrm{days}$ and $0.2 \ M_{\mathrm{Jup}}< M_p \sin i < 5 \ M_{\mathrm{Jup}}$.) We also plot the $\rho_p-R_p$ regions for two candidates, one with strong physical plausibility and one with weak plausibility.

A quantitative metric is necessary for comparing a given candidate's allowed $\rho_p-R_p$ domain to the region occupied by known hot Jupiters. We first calculated a kernel density estimation of the values for known hot Jupiters in $\ln\rho_p$-$R_p$ space, denoted $f(R_p, \ln\rho_p)$ and illustrated with the gray gradient in Figure~\ref{rho vs R plot}. We then calculated the integral, $S$, of the kernel density over each candidate's allowed domain,
\begin{equation}
S = \iint_D f(R_p, \ln\rho_p) dR_p d\ln\rho_p,
\end{equation}
where 
\begin{equation}
D = \{(R_p, \ln\rho_p):A_g \in [0.1, 0.3], i \in [15^{\circ},85^{\circ}]\}.
\end{equation}

We calculated the $S$ integrals for all candidates with probabilities, $P > 0.5$. We then removed all candidates with $S < 5$. Though the choice of this threshold is somewhat arbitrary, it corresponds to $\gtrsim 50 \%$ of a candidate's $\rho_p$-$R_p$ domain overlapping with that of the known planet detections.\footnote{It is important to note that this is a quite conservative vetting procedure, since the albedo and inclination ranges are not as wide as possible, and since there could feasibly be new planet detections on the outskirts of the known domain. However, in the process of selecting the best possible candidates, conservative vetting is a logical solution.} We also verified that the four recovered transiting hot Jupiters from Section~\ref{testing on transiting data} had $S$ integral measures greater than this threshold. 

\begin{figure}
\epsscale{1.3}
\plotone{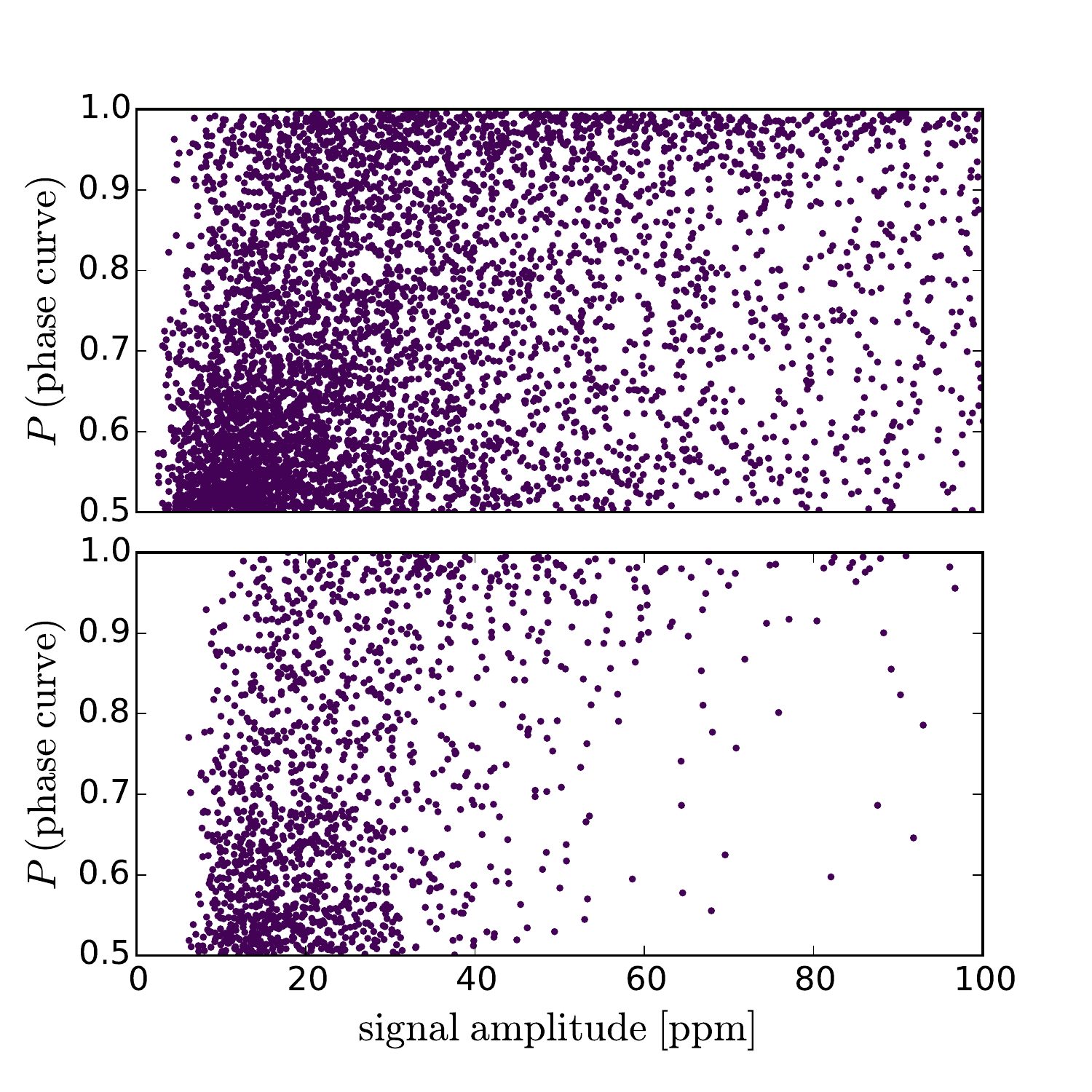}
\caption{Top panel: probability versus amplitude for the candidates with $P>0.5$ (top half of Figure~\ref{real data prediction}). Bottom panel: remaining candidates after the filters from Sections~\ref{binary star filtering} and~\ref{candidate vetting} were applied.}
\label{before and after filters}
\end{figure}

Figure~\ref{before and after filters} shows the results of the binary star filtering (Section~\ref{binary star filtering}) and the physical plausibility vetting (Section~\ref{candidate vetting}) on the aggregate of candidates with $P>0.5$. These filters clearly have a dramatic effect on the number of plausible planetary phase curve candidates.

\section{Results: A Catalog of Candidate Planets with Phase Curves}
\label{candidate catalog}

In the remainder of the paper, we focus our analysis on the subset of highest probability candidates. Figures~\ref{synthetic data prediction} and~\ref{transiting data prediction} show that true phase curves should be reported with probabilities very close to 1. To this end, we selected all candidates with $P > 0.97$ and $\mathrm{Amp} < 70$ ppm. Lastly, we established a final candidate list through a small amount of visual inspection. In particular, we inspected the Lomb-Scargle periodograms and removed any that showed signs of significant peaks apart from the primary peak. We also examined the $2P$-folded phase curves and checked for any subtle indications of amplitude inconsistency in each half.

Following this final vetting procedure, we present 60 remaining high probability candidates. In Table~\ref{candidate table}, we list their parameters. The parameters $K_p$,  $T_{\mathrm{eff}}$, $R_{\star}$, $M_{\star}$, and Fe/H are from the Q1-Q17 DR 25 \textit{Kepler} stellar catalog \citep{2017ApJS..229...30M}. The uncertainties on the periods were calculated using a Gaussian fit to the peak in the Lomb-Scargle periodogram. The units of $A_{\mathrm{refl}}$ and $A_{\mathrm{ellip}}$ are ppm.

We also report estimates of the candidates' minimum RV semi-amplitudes, $K\sin i$. To calculate this, we note that the amplitude of the ellipsoidal variation component, $A_{\mathrm{ellip}}$, obtained via equation~\ref{ellipsoidal variation}, is directly related to $K$ through the relation,
\begin{equation}
K \sin i = \frac{A_{\mathrm{ellip}}}{\alpha_2 }\left(\frac{G M_{\star}}{a} \right)^{1/2} \left(\frac{a}{R_{\star}}\right)^3
\end{equation}
The errorbars on $K \sin i$ were obtained via propagation of the uncertainties on the KIC stellar parameters and the $A_{\mathrm{ellip}}$ estimates obtained from the least-squares fit introduced in the first paragraph of Section~\ref{candidate vetting}. Unfortunately these uncertainties are quite large, mostly due to the lack of constraint on the stellar parameters.

An online repository of the candidates may be found at \href{https://smillholland.github.io/Non-transiting_HJs/}{https://smillholland.github.io/Non-transiting\_HJs/}. The repository contains a downloadable candidate catalog and a variety of diagnostic plots, including the candidate phase curves, periodograms, and $\rho_p-R_p$ diagrams. We also provide information relevant to follow-up observations, including estimates of the orbital ephemerides.

\begin{table*}[t]
\caption{A catalog of 60 high probability non-transiting hot Jupiter candidates} \label{candidate table} 
\begin{center}
\begin{longtable}{c c c c c c c c c c c}\hline\hline  KIC & $K_p$ & $T_{\mathrm{eff}} \ [\mathrm{K}]$ & $R_{\star} \ [R_{\odot}]$ & $M_{\star} \ [M_{\odot}]$ & Fe/H & $P$ [days] & $A_{\mathrm{refl}}$ & $A_{\mathrm{ellip}}$ & $K\sin i \ [\mathrm{m/s}]$ & Prob \\\hline \begin{filecontents*}{Candidate_IDs_061017_prob_97_amp_70ppm_Final_List_Part1.csv}
KepID, RA, Dec, Teff, Rs, Ms, FeH, Kp, period, period_unc, Ksini, Ksini_unc, T_0_ellip, T_0_refl, A_refl, A_ellip, phase_offsets, SNR, probability
2706947,290.01038,37.925758,6771,1.29,1.15,-0.44,13.7,2.5762,0.0018,150,144,2456423.956,2456424.276,48,3.9,0.124,0.7,0.981
2708787,290.44894,37.988708,4243,0.63,0.62,-0.12,15.1,1.8296,0.0008,326,188,2456423.177,2456423.266,49.6,6,0.048,0.3,0.987
3217078,285.78049,38.345821,4780,0.77,0.75,0.14,15.7,1.853,0.0008,389,200,2456423.703,2456423.99,50.7,9.2,0.155,0.24,0.984
3347307,292.73236,38.442211,5955,0.91,1,-0.16,14.6,1.8329,0.0008,251,248,2456425.179,2456425.039,31.5,5.7,-0.076,0.3,0.992
3539728,290.04871,38.67337,6075,0.97,1.05,-0.12,15.3,1.6002,0.0006,257,268,2456424.551,2456424.227,42.9,8.1,-0.203,0.29,0.982
3964318,294.08643,39.041279,4544,0.64,0.71,-0.08,14.1,1.7412,0.0007,133,108,2456424.157,2456424.298,28,2.3,0.081,0.32,0.977
4753174,293.64124,39.873501,6592,1.19,1.21,-0.12,14.3,1.685,0.0008,55,63,2456424.284,2456424.244,47.4,2.2,-0.024,0.6,0.981
4914087,288.88971,40.089748,5505,0.88,0.83,-0.18,15,2.2909,0.0017,496,372,2456425.416,2456424.068,22.8,9.8,-0.089,0.21,0.984
5001685,288.89331,40.110828,5713,1.01,1.01,0.18,14,2.2906,0.0014,221,108,2456424.198,2456424.058,15.1,4.8,-0.061,0.26,0.991
5354490,289.34195,40.560989,5638,0.97,1,0.21,14.7,2.6102,0.0016,130,112,2456425.248,2456425.502,47.1,2.1,0.097,0.52,0.992
5435675,287.57367,40.673309,6032,0.97,1.05,-0.06,15.5,1.5229,0.0006,45,85,2456423.69,2456424.99,29.5,1.6,-0.146,0.2,0.987
5479689,299.3256,40.63822,5360,0.91,0.92,0.24,13.8,1.7012,0.0008,158,111,2456422.53,2456422.769,15.5,5,0.141,0.24,0.987
5566612,299.59052,40.70116,5799,1.1,0.97,0,14.9,2.7099,0.002,513,454,2456423.38,2456423.217,27.1,11.4,-0.06,0.2,0.972
5597644,283.26529,40.809235,4608,0.67,0.66,-0.22,14.9,1.6228,0.0007,473,170,2456390.253,2456390.364,38.4,11.3,0.068,0.36,0.985
5716330,294.98325,40.97084,5952,1.3,1.05,0,13.9,1.5689,0.0006,57,32,2456424.823,2456424.688,29.8,4.6,-0.086,0.5,0.996
5781247,288.71573,41.030861,5561,0.85,0.97,0.07,15.6,1.6547,0.0008,59,112,2456423.404,2456423.32,37.1,1.5,-0.051,0.26,0.974
5878307,292.46521,41.16589,6856,1.42,1.28,-0.26,13.6,2.0466,0.0013,116,112,2456423.898,2456423.993,18.9,5,0.047,0.37,0.978
6047853,293.48605,41.320229,6422,1.42,1.22,-0.06,13.7,3.4193,0.0029,129,83,2456421.478,2456421.436,23.3,2.8,-0.012,0.44,0.985
6065597,297.93222,41.363781,5726,1.18,0.92,-0.08,13.7,3.9872,0.0045,254,232,2456427.461,2456427.243,20.4,4,-0.055,0.35,0.987
6388466,298.30063,41.765461,5988,1.07,0.99,-0.12,14.5,1.269,0.0004,64,64,2456423.839,2456423.874,43.2,4.4,0.028,0.47,0.993
6603087,290.87234,42.057789,5522,0.87,0.97,0.14,15.4,1.3457,0.0005,50,82,2456424.07,2456423.882,35.3,1.9,-0.14,0.26,0.975
6613542,293.95676,42.04937,6113,0.98,0.96,-0.34,14.1,4.4023,0.005,240,269,2456423.365,2456421.787,20.8,1.6,0.141,0.31,0.976
6675953,286.60907,42.173359,5565,0.82,0.8,-0.36,13.7,2.9651,0.0023,135,118,2456425.812,2456425.424,20.4,1.4,-0.131,0.4,0.988
6783562,293.62296,42.204281,6930,1.19,1.22,-0.32,13.6,2.0883,0.0014,243,267,2456423.774,2456425.532,42.7,6.3,-0.158,0.84,0.992
6976754,298.73401,42.435349,6607,1.2,1.22,-0.1,14.5,1.4196,0.0005,64,73,2456422.823,2456422.912,56.6,3.5,0.063,0.57,0.989
7045031,295.09708,42.596489,5445,0.8,0.78,-0.34,15.8,1.2886,0.0007,44,76,2456423.447,2456423.386,52.3,1.9,-0.047,0.3,0.976
7512130,286.39355,43.137791,6108,1.03,1.13,0.07,14.3,1.8712,0.0008,47,64,2456423.101,2456423.327,44.7,1.2,0.121,0.56,0.996
7596734,287.97717,43.257469,6078,0.88,0.96,-0.38,15,2.5893,0.0016,230,264,2456423.931,2456424.504,21.6,2.8,0.221,0.2,0.989
7735171,283.90421,43.448952,5738,0.95,0.97,0.02,15.2,1.8334,0.0008,186,190,2456424.787,2456424.537,37.2,5.1,-0.136,0.24,0.984
7938689,282.07098,43.784019,5269,0.92,0.83,0.07,15.4,1.841,0.0009,148,125,2456423.746,2456424.111,48.6,4.9,0.199,0.3,0.991
7978458,297.56415,43.761002,6188,1.08,1.03,-0.14,15,1.4349,0.0005,77,82,2456423.95,2456424.173,29.2,4,0.155,0.27,0.986
8026887,291.12845,43.874939,5988,1.22,1.01,-0.08,13.9,1.9232,0.0009,234,113,2456423.311,2456423.353,35.1,11.3,0.022,0.54,0.995
8042004,296.18439,43.843666,6223,1.64,1.3,0.14,13.3,1.9498,0.0011,42,39,2456423.806,2456425.551,14.5,3.3,-0.105,0.41,0.991
8098079,291.98013,43.904411,6198,1,1.07,-0.16,15.1,1.2873,0.0004,89,106,2456423.051,2456424.142,34.4,4.3,-0.152,0.27,0.992
8122501,299.10883,43.985931,5543,0.82,0.73,-0.54,14.7,1.029,0.0003,98,79,2456424.044,2456423.926,64,6.9,-0.114,0.61,0.98
8121913,298.9678,43.909389,5702,1.45,1.03,0.14,11.7,3.2943,0.0029,61,66,2456422.928,2456425.762,18.5,2.1,-0.14,0.56,0.988
8358253,289.07788,44.39481,5985,0.85,0.94,-0.36,14.5,1.5602,0.0006,609,541,2456423.669,2456423.483,33.7,16,-0.119,0.4,0.996
8364428,291.38873,44.352821,6525,1.59,1.05,-0.56,13.4,1.5737,0.0008,96,105,2456424.505,2456424.61,49.9,12.6,0.067,0.96,0.985
8389838,299.19931,44.326302,5725,0.94,0.87,-0.24,15,2.0092,0.0011,226,204,2456423.97,2456424.067,23.5,6,0.548,0.18,0.972
8410490,282.08298,44.435329,5821,0.87,0.91,-0.28,14.1,1.3621,0.0004,88,83,2456424.016,2456424.105,23.3,3.4,0.066,0.32,0.994
8499974,293.68069,44.573639,5952,0.84,0.93,-0.36,14.9,1.4916,0.0005,238,222,2456424.373,2456424.174,35.9,6.6,-0.133,0.38,0.981
8559208,291.49591,44.600159,6029,0.82,0.92,-0.5,14.5,2.1802,0.0012,368,338,2456423.193,2456425.251,25.9,5.1,-0.056,0.3,0.992
8752590,291.47281,44.956589,5750,0.95,0.97,0,15.6,1.088,0.0003,99,100,2456423.849,2456423.908,58.4,6.6,0.054,0.35,0.98
8832613,296.77988,45.06234,4571,0.66,0.67,-0.2,15.8,1.5788,0.0008,393,187,2456424.156,2456424.415,53.8,9.1,0.164,0.27,0.971
8847921,300.94221,45.09841,5739,1.14,0.93,-0.06,14.7,1.7975,0.0008,32,46,2456423.199,2456423.33,28,1.7,0.073,0.23,0.985
\end{filecontents*}
\csvreader[late after line=\\]
{Candidate_IDs_061017_prob_97_amp_70ppm_Final_List_Part1.csv}{KepID=\KIC, RA = \RA, Dec = \Dec, Teff=\Teff, Rs=\Rstar, Ms=\Mstar, FeH=\FeH, Kp=\Kp, period=\period, period_unc = \periodUnc, Ksini = \Ksini, Ksini_unc = \KsiniUnc, T_0_ellip = \Tellip, T_0_refl = \Trefl, A_refl=\Arefl, A_ellip=\Aellip, phase_offsets=\phaseoffsets, SNR=\SNR, probability=\probability}
{\KIC & \Kp & \Teff & \Rstar & \Mstar & \FeH  & {\period$\pm$\periodUnc} & \Arefl & \Aellip & {\Ksini$\pm$\KsiniUnc} & \probability }
\end{longtable}
\end{center}
\end{table*}

\begin{table*}[t]
\caption{A catalog of 60 high probability non-transiting hot Jupiter candidates} \label{candidate table} 
\begin{center}
\begin{longtable}{c c c c c c c c c c c}\hline\hline  KIC & $K_p$ & $T_{\mathrm{eff}} \ [\mathrm{K}]$ & $R_{\star} \ [R_{\odot}]$ & $M_{\star} \ [M_{\odot}]$ & Fe/H & $P$ [days] & $A_{\mathrm{refl}}$ & $A_{\mathrm{ellip}}$ & $K\sin i \ [\mathrm{m/s}]$ & Prob \\\hline \begin{filecontents*}{Candidate_IDs_061017_prob_97_amp_70ppm_Final_List_Part2.csv}
KepID, RA, Dec, Teff, Rs, Ms, FeH, Kp, period, period_unc, Ksini, Ksini_unc, T_0_ellip, T_0_refl, A_refl, A_ellip, phase_offsets, SNR, probability
8885337,291.7265895,45.12058,6950,1.39,1.43,0.07,13.9,2.2199,0.0013,139,171,2456425.223,2456423.863,37.8,4.2,-0.113,0.55,0.978
9040864,297.6082695,45.31879,5774,1.13,0.92,-0.12,14.7,2.1063,0.0013,81,90,2456423.122,2456423.45,46.2,3.2,0.156,0.36,0.971
9287646,294.47601,45.790329,6258,1.03,1.09,-0.16,14.1,2.5591,0.0018,568,569,2456423.235,2456425.644,18.9,9,-0.059,0.27,0.986
9475045,295.65652,46.07238,6143,1.12,1.19,0.21,15.1,1.173,0.0003,51,63,2456423.773,2456424.707,38.4,3.6,-0.204,0.33,0.995
9570402,282.57468,46.246761,6103,0.95,1.03,-0.2,14,1.683,0.0007,156,151,2456425.156,2456424.078,48.1,4.3,-0.141,0.7,0.994
9724972,297.30585,46.40382,6017,0.99,0.91,-0.38,14.2,2.0872,0.0013,256,227,2456424.044,2456424.244,32.6,6.7,0.596,0.47,0.978
10068024,288.84711,47.022911,6349,1.06,1.08,-0.22,13.1,2.0735,0.0012,229,243,2456423.228,2456423.072,20.6,5.6,-0.075,0.47,0.985
10091175,297.48868,47.075291,6063,1.01,1.12,0.07,14.5,1.8757,0.0009,394,385,2456424.557,2456424.704,39.4,9.9,0.578,0.42,0.991
10619862,298.57874,47.862099,5783,0.88,0.98,-0.08,14.5,3.0538,0.0025,190,247,2456422.91,2456422.961,33.7,1.8,0.017,0.38,0.988
10931452,296.54721,48.30257,6666,1.54,1.14,-0.38,12.7,2.7001,0.0023,108,91,2456422.603,2456422.62,21.7,4.5,0.006,0.66,0.971
11152428,298.03064,48.79137,6774,1.35,1.25,-0.26,13.8,1.1726,0.0003,89,85,2456424.068,2456423.968,33.6,8.8,-0.585,0.5,0.982
11235536,287.17719,48.932812,5631,0.81,0.88,-0.24,15.5,1.4585,0.0005,307,262,2456424.454,2456424.454,54.7,8.8,0,0.35,0.992
11362225,297.0674805,49.18819,6623,1.9,1.45,0.04,12,2.7513,0.0021,103,71,2456425.321,2456422.952,42.4,5.9,0.139,1,0.995
12207153,290.57666,50.807789,5569,0.89,1.01,0.21,14.8,1.4225,0.0005,257,217,2456422.803,2456423.086,30.2,8.6,0.699,0.24,0.979
12357100,291.47293,51.111431,5880,0.92,1.01,-0.06,15.1,2.6613,0.0021,258,302,2456423.005,2456423.018,36.7,3.2,0.005,0.25,0.979
\end{filecontents*}
\csvreader[late after line=\\]
{Candidate_IDs_061017_prob_97_amp_70ppm_Final_List_Part2.csv}{KepID=\KIC, RA = \RA, Dec = \Dec, Teff=\Teff, Rs=\Rstar, Ms=\Mstar, FeH=\FeH, Kp=\Kp, period=\period, period_unc = \periodUnc, Ksini = \Ksini, Ksini_unc = \KsiniUnc, T_0_ellip = \Tellip, T_0_refl = \Trefl, A_refl=\Arefl, A_ellip=\Aellip, phase_offsets=\phaseoffsets, SNR=\SNR, probability=\probability}
{\KIC & \Kp & \Teff & \Rstar & \Mstar & \FeH  & {\period$\pm$\periodUnc} & \Arefl & \Aellip & {\Ksini$\pm$\KsiniUnc} & \probability} 
\end{longtable}
\end{center}
\end{table*}

\subsection{Albedo constraints}
\label{albedo constraints}
Under the assumption that the majority of remaining high probability candidates correspond to true planets, we are presented with a collection of phase curves that may be examined in greater detail. One quantity of interest is the planetary albedo. Without additional information, it is impossible to disentangle the degeneracy between the albedo, planet radius, and inclination. However, we can consider the composite quantity,  
$A_g {R_p}^2 \sin i$, which can be calculated directly from the observable quantities as follows:

\begin{equation}
A_g {R_p}^2 \sin i = 2 A_{\mathrm{refl}} \left(\frac{G M_{\star}}{4\pi^2}\right)^{2/3} P^{4/3}.
\end{equation}
Here, $P$ and $A_{\mathrm{refl}}$ are the period and reflection semi-amplitude of the candidate phase curve, and $M_{\star}$ is taken from the \textit{Kepler} stellar catalog.

\begin{figure}
\epsscale{1.25}
\plotone{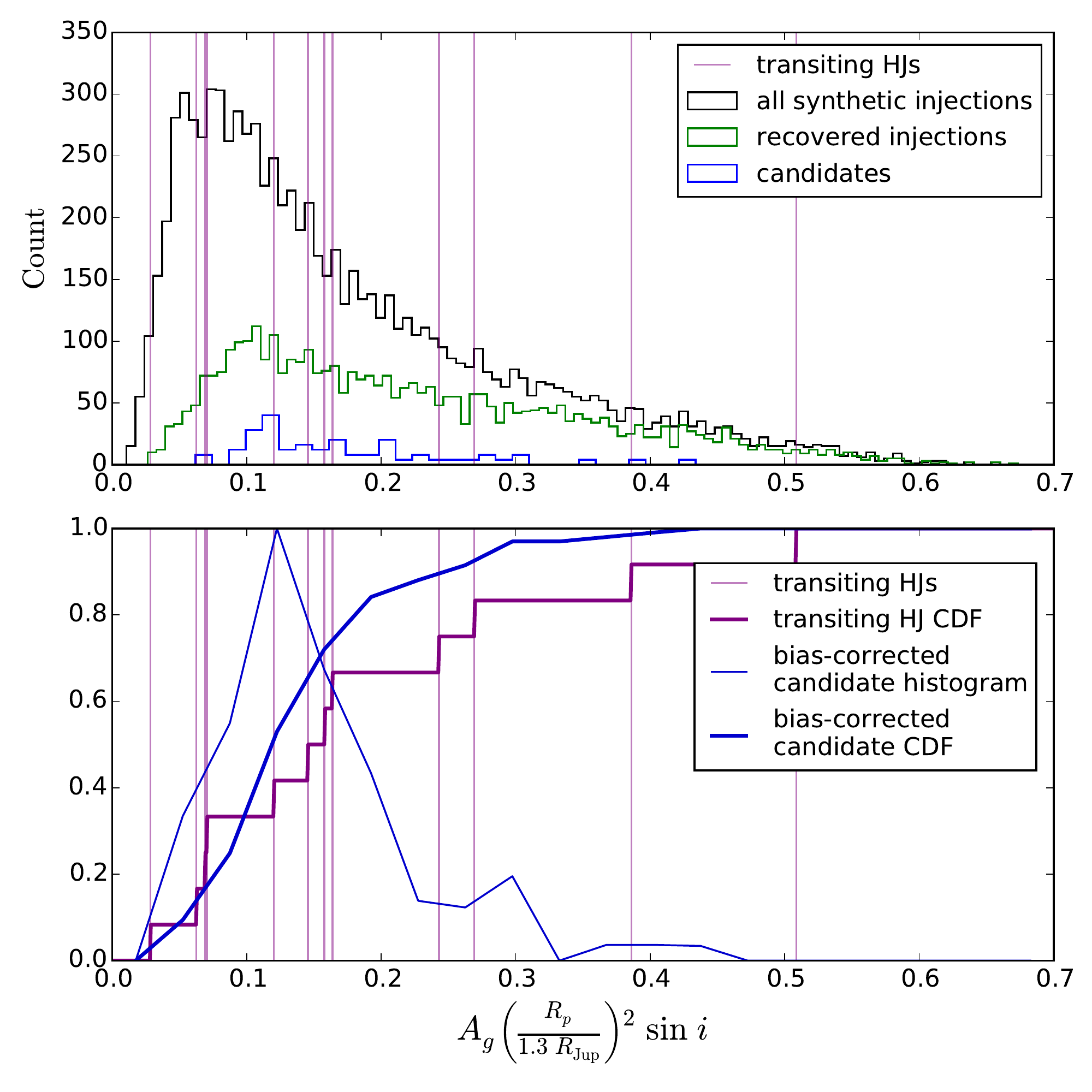}
\caption{Constraints on the candidates' albedo through the quantity $A_g {\left(\frac{R_p}{1.3 \ R_{\mathrm{Jup}}} \right)}^2 \sin i$. The top panel contains histograms for the injected and recovered synthetic data and the candidates, where the candidate histogram has been normalized to be on the same scale as the synthetic data histograms. The bottom panel contains the bias-corrected candidate histogram and cumulative distribution function. Both panels show the values for known transiting planets with phase curves from \cite{2015ApJ...804..150E}, and the bottom panel shows the corresponding CDFs.}
\label{albedo distribution}
\end{figure}

In Figure~\ref{albedo distribution}, we display the results of the $A_g {R_p}^2 \sin i$ calculations. In the top panel, we show the histogram of $A_g {R_p}^2 \sin i$ for the candidates in blue. We also show the histograms of all synthetic injections and the recovered synthetic injections from Section~\ref{synthetic injection and recovery}. The ratio of these two gives a recovery efficiency curve, and in the bottom panel, we use this efficiency curve to construct the bias-corrected histogram for the candidates. We compare these results to 12 known transiting hot Jupiters with phase curves from \cite{2015ApJ...804..150E}. It is clear from the comparison of the bias-corrected candidate CDF to the transiting hot Jupiter CDF that the candidates are a good match to known planets. The candidate CDF is slightly shifted towards lower values, which is an expected outcome of the bias correction. Furthermore, the bias-corrected histogram peaks at $A_g {R_p}^2 \sin i \sim 0.1$, providing further evidence that hot Jupiters are generally fairly dark.

\begin{figure}
\epsscale{1.25}
\plotone{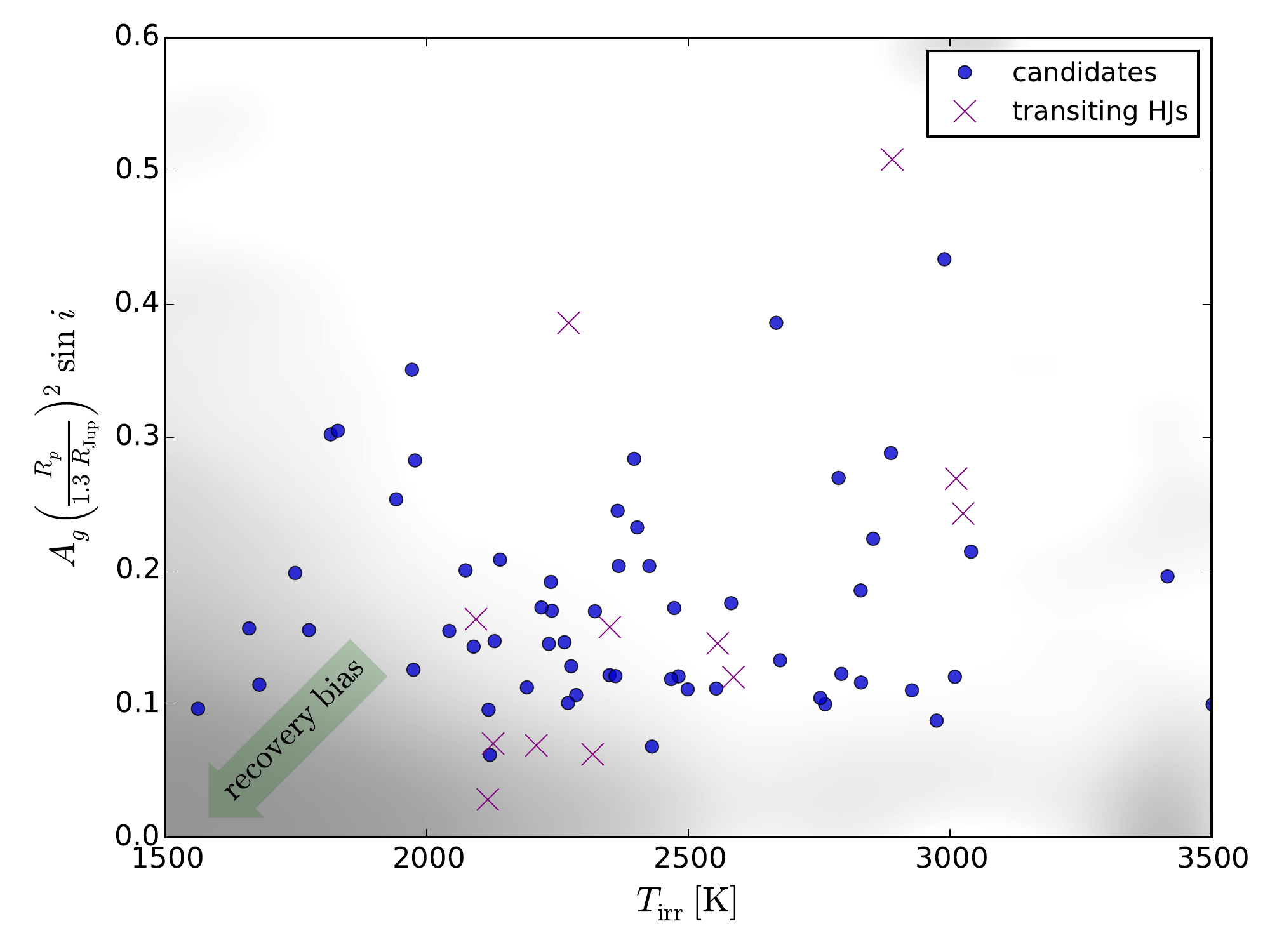}
\caption{$A_g {\left(\frac{R_p}{1.3 \ R_{\mathrm{Jup}}} \right)}^2 \sin i$ versus $T_{\mathrm{irr}}$ for the candidates in blue and for transiting hot Jupiters from \cite{2015ApJ...804..150E} in purple. The gray gradient shows the region that is biased against detections because of low albedos and/or planet radii and smaller incident flux. }
\label{Ag_Rsq_sini_vs_T_irr}
\end{figure}

It is interesting to check whether the candidates show any systematic trends in albedo with other quantities. In Figure~\ref{Ag_Rsq_sini_vs_T_irr}, we plot $A_g {R_p}^2 \sin i$ vs. $T_{\mathrm{irr}}$, where $T_{\mathrm{irr}}$ is defined as
\begin{equation}
\label{Tirr}
T_{\mathrm{irr}} = T_{\mathrm{eff}}\left(\frac{R_{\star}}{a}\right)^{1/2},
\end{equation}
such that the incident stellar irradiation is $F_0 = \sigma T_{\mathrm{irr}}^4$ \citep{2012ApJ...751...59P}. $T_{\mathrm{irr}}$ is not the planet's equilibrium temperature, which is rather given by
\begin{equation}
T_{\mathrm{eq}} = T_{\mathrm{eff}}\left(\frac{R_{\star}}{a}\right)^{1/2}\left[f (1-A_B)\right]^{1/4},
\end{equation}
where $A_B$ is the Bond albedo and  $f$ is a re-radiation factor ranging between 1/4 (homogeneous redistribution) and 2/3 (instant re-radiation) \citep{2013ApJ...772...51E}. However, $T_{\mathrm{irr}}$ is a close proxy to $T_{\mathrm{eq}}$. 

The gray gradient in Figure~\ref{Ag_Rsq_sini_vs_T_irr} illustrates regions with stronger detection bias. This was calculated by finding the point density of recovered synthetic injections and dividing it by the point density of all synthetic injections.

There is little obvious trend in $A_g {R_p}^2 \sin i$ vs. $T_{\mathrm{irr}}$. There is a weak indication that the average $A_g {R_p}^2 \sin i$ decreases as a function of $T_{\mathrm{irr}}$ up to $T_{\mathrm{irr}} \sim 2700 \ \mathrm{K}$, at which point the phase curve would start involving substantial thermal radiation, and $A_g$ would not just include reflected light. This may suggest that cooler planets are brighter on average, with a greater proportion of reflective clouds. It is not surprising to see the lack of a strong relationship, however. We are using \textit{Kepler} stellar catalog parameters, so $T_{\mathrm{irr}}$ is not known to high accuracy. It would be valuable to revisit this after more precise stellar parameters have been obtained. Data from the \textit{Gaia} spacecraft \citep{2016A&A...595A...1G} may be a near-term resource for these improved estimates. \\ \\

\subsection{Offsets of the phase curve maxima}

A phase offset between the phase curve maximum and sub-stellar point has been observed for most planetary phase curves. The maximum is shifted eastward for thermal phase curves \citep{2002AA...385..166S, 2007Natur.447..183K}, as superrotating jets advect the hottest region downwind. Both eastward and westward shifts have been observed for optical phase curves \citep{2015AJ....150..112S, 2015ApJ...804..150E}. 

For phase curves of transiting planets, the phase offset is easily determined by comparing the phase curve maximum to the secondary eclipse. Determining $\Delta \theta$ is not easy for non-transiting planets, but it is possible in some cases. If beaming and/or ellipsoidal components of the light curve are detected, the phase offset can be obtained from the phase relationship between the reflected light and beaming/ellipsoidal components.

Towards this goal, we used the three-component phase curve fits (equation~\ref{three component model}) of the candidates from Section~\ref{candidate vetting}. We calculated the phase offset between the reflection maximum and sub-stellar point using $\Delta \theta = \theta_{\mathrm{refl}} - \theta_{\mathrm{ellip}}$. For the candidate pictured in Figure~\ref{three component fit}, the phase offset is consistent with an eastward shift of the phase curve maximum.

\begin{figure}
\epsscale{1.25}
\plotone{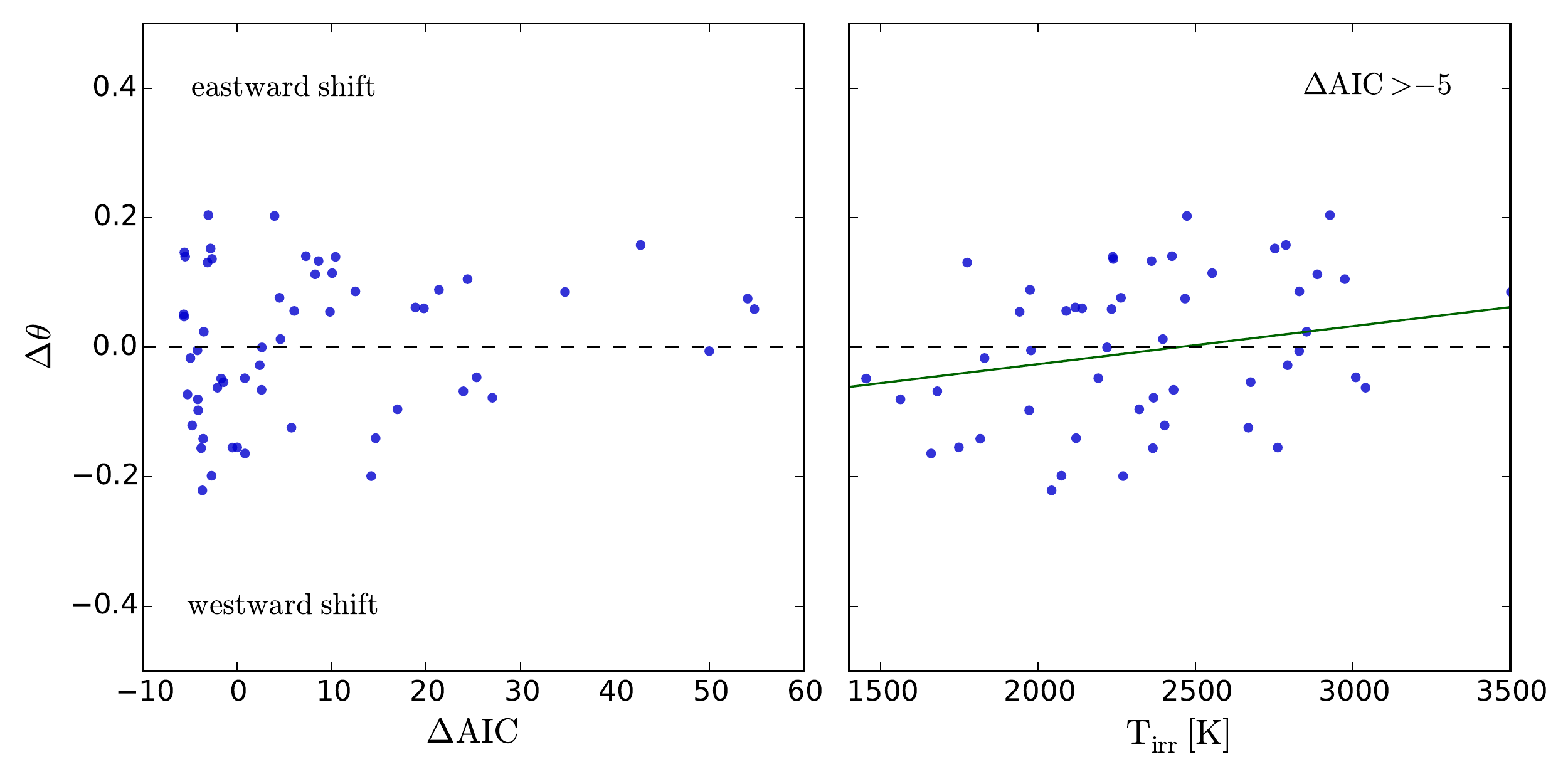}
\caption{The phase offset $\Delta \theta = \theta_{\mathrm{refl}} - \theta_{\mathrm{ellip}}$ versus $\Delta \mathrm{AIC}$ and $T_{\mathrm{irr}}$. $\Delta \mathrm{AIC}$ corresponds to the difference in $\mathrm{AIC}$ between the one-component and three-component phase curve models. The larger the difference, the more the three-component model is favored. In the right panel, the candidates have been restricted to those with $\Delta \mathrm{AIC} > -5$, and a least-squares line has been overplotted in green. $\Delta \theta$ exhibits a positive correlation with both $\Delta \mathrm{AIC}$ and $T_{\mathrm{irr}}$.}
\label{phase offset vs AIC and Tirr}
\end{figure}

For each of the candidate phase curve fits, we also calculated the differences in the Aikaike Information Criterion (AIC) between the one-component and three-component fits. If the three-component fit is favored, then $\Delta \mathrm{AIC}$ is large and the ellipsoidal/beaming components are more significant. In Figure~\ref{phase offset vs AIC and Tirr}, we plot $\Delta \theta$ versus $\Delta \mathrm{AIC}$ and $T_{\mathrm{irr}}$. $\Delta \theta > 0$ corresponds to eastward shifts in the phase curve maxima (i.e. the phase curve peaks before the substellar point), while $\Delta \theta < 0$ corresponds to westward shifts.

From the left panel in Figure~\ref{phase offset vs AIC and Tirr}, we see that candidates with smaller $\Delta \mathrm{AIC}$ favor westward shifts in the phase curve maxima, whereas the maxima of candidates with larger $\Delta \mathrm{AIC}$ tend to be eastward shifted. Moreover, the right panel provides weak evidence for a positive correlation of $\Delta \theta$ with $T_{\mathrm{irr}}$. Both of these correlations strengthen appreciably when considering candidates with higher probability thresholds (e.g. $P>0.985$, rather than the 0.97 used here). We emphasize again that we are using \textit{Kepler} stellar catalog parameters in the calculation of $T_{\mathrm{irr}}$, which likely explains much of the scatter in Figure~\ref{phase offset vs AIC and Tirr}. It will be valuable to see how the results improve with more reliable spectroscopic and astroseismic stellar parameter estimates.

It is logical that $\Delta \mathrm{AIC}$ and $T_{\mathrm{irr}}$ should show a similarly signed correlation with $\Delta \theta$, as $\Delta \mathrm{AIC}$ is closely related to $T_{\mathrm{irr}}$. As $\Delta \mathrm{AIC}$ increases, the ellipsoidal component of the light curve becomes more significant. Among other dependencies, this correlates with a small $a$ or large $R_{\star}$ (see equation~\ref{ellipsoidal variation}), both of which would act to increase $T_{\mathrm{irr}}$ (equation~\ref{Tirr}).   

Several previous studies have also observed a positive trend in $\Delta \theta$ vs. $T_{\mathrm{irr}}$ or $T_{\mathrm{eq}}$ \citep{2015PASP..127.1113A, 2015AJ....150..112S, 2015ApJ...804..150E, 2017arXiv170300496S}. Optical phase curves contain contributions from both reflected light and thermal radiation. For the hottest planets, the phase curve contains a significant thermal contribution, and the eastward phase shift matches that which is observed in IR phase curves. This phase shift is thought to be due to a superrotating equatorial jet advecting the hot spot eastward of the substellar point \citep[e.g.][]{2002AA...385..166S}. For the coolest planets, the phase curve is reflection-dominated, and the westward phase shift has been attributed to the presence of reflective clouds condensing in the cooler regions of the planet, westward of the substellar point \citep[e.g.][]{2017arXiv170300496S}. The linear correlation in $\Delta \theta$ vs. $T_{\mathrm{irr}}$ could then just arise from the relative contribution of each phase curve component.

The presence of both thermal and reflection components in the phase curve makes it challenging to draw physical interpretations of the atmospheric dynamics. Considering the hotspot alone, an increase in $T_{\mathrm{irr}}$ should result in a \textit{decrease} in the magnitude of the phase offset, as heat redistribution becomes less efficient \citep{2012ApJ...751...59P}. This arises from the balance between radiative and advective timescales, $t_{\mathrm{rad}}$ and $t_{\mathrm{adv}}$. With 
\begin{equation}
t_{\mathrm{rad}} = \frac{c_P P}{\sigma_{\mathrm{SB}} \ g_p \ {T_{\mathrm{max}}}^3}
\end{equation}
and 
\begin{equation}
t_{\mathrm{adv}} = \frac{R}{v_{\theta}},
\end{equation}
$t_{\mathrm{adv}}/t_{\mathrm{rad}}$ increases rapidly with $T_{\mathrm{irr}}$, such that the heat redistribution efficiency and the hotspot offset both decrease with $T_{\mathrm{irr}}$ \citep{2012ApJ...751...59P}. 

Ohmic dissipation, which becomes significant at large $T_{\mathrm{irr}}$ due to an increase in thermal ionization and the resulting electrical conductivity, should also act to slow the zonal winds and reduce the magnitude of the hotspot offset \citep{2010ApJ...719.1421P, 2010ApJ...724..313P,2013ApJ...776...53B}. This effect works constructively with the balance between $t_{\mathrm{adv}}$ and $t_{\mathrm{rad}}$ described above. 

In order to see these theorized trends of $\Delta \theta$ vs. $T_{\mathrm{irr}}$ in the data, it would be necessary to isolate the thermal component from the reflection component of the phase curve \citep[e.g.][]{2016NatAs...1E...4A}. This would be a valuable undertaking once the candidates presented in this paper are confirmed and the planets' irradiation temperatures may be determined more accurately.

\section{Discussion}

\subsection{Caveats \& Future Work}

There is an important caveat related to the identification and characterization of non-transiting planet candidates presented in this paper. The concern relates to the effects of nearby stellar companions.  Using high-resolution optical and near-IR imaging, \cite{2017AJ....153...71F} found that $\sim 30\%$ of KOI host stars have at least one companion star within 4\arcsec. Since the \textit{Kepler} detector has $\sim$4\arcsec wide pixels and the aperture photometry is obtained from at least a few pixels, this implies that $\gtrsim 30\%$ of the candidates presented here will have signal amplitudes that are smaller than they would otherwise be without nearby companions.

In addition to influencing the constraints on the candidates' albedo, planet radius, and inclination (see Section~\ref{albedo constraints}), this photometric contamination can also result in false positive phase curve detections in the form of background or foreground binary star phase curves, analogous to the dominant transit false positive sources \citep[e.g.][]{2016ApJ...822...86M}. Although the binary star filtering and physical plausibility vetting outlined in Sections~\ref{binary star filtering} and~\ref{candidate vetting} have certainly helped reduce some of these false positives, we do not claim to have fully accounted for the problem.

It might be possible to further reduce the effects of background or foreground binary stars by including them in the synthetic dataset that was first produced in Section~\ref{light curve injection}. To do this, one could first assume a distribution of angular separations and magnitude differences of nearby stellar companions using the results of \cite{2017AJ....153...71F}. Then, the distribution of short-period binary stars could be empirically modeled after the findings of \cite{2016AJ....151...68K} and other papers in this series. After injecting signals appropriate to these distributions, one could search for predictors to add to the logistic regression that would help distinguish the binary signals from the planetary phase curve signals.

\subsection{Candidate follow-up observations}

The main result of this paper is a catalog of 60 non-transiting, short-period giant planet candidates with phase curves (presented in Section~\ref{candidate catalog}). These candidates can be confirmed with follow-up radial velocity observations. Even though the target stars are dim, hot Jupiters are readily detectable even with low RV precision. Moreover, the known orbital ephemerides from the phase curves will significantly aid the RV measurements.  Once the RVs confirm a planet candidate's existence, the comparison between the RV and photometric ephemerides will allow for a more precise measurement of the shift between the phase curve maximum and sub-stellar point. Moreover, the combined RV-photometric modeling could permit a break in the mass/inclination degeneracy.

After RV confirmation, other valuable follow-up observations include \textit{Spitzer} infrared photometry of the brightest targets. There are very few planets with both optical and infrared phase curve observations \citep{2010ApJ...710...97C, 2013ApJ...776L..25D}. The combination of these observations, and particularly a comparison of the offset of the phase curve maxima, would be strongly constraining on the planets' atmospheric dynamics. \\

\subsection{Candidate stellar metallicities and effective temperatures}

It is well-known that the occurrence rate of giant planets correlates strongly with stellar metallicity \citep{2004AA...415.1153S, 2005ApJ...622.1102F, 2010PASP..122..905J}. Consideration of the stellar metallicity of the highest probability candidates thus provides a strong independent check on their validity; the distribution of candidate metallicities should be systematically higher than the distribution of metallicities of the rest of the population. 

The metallicities reported in the \textit{Kepler} stellar catalog have been primarily photometrically-derived \citep{2011AJ....142..112B} and have also been shown to systematically underestimate the true metallicity and scatter \citep{2014ApJ...789L...3D, 2017arXiv170310400P}. 
\cite{2014ApJ...789L...3D} used LAMOST spectroscopic data to show that the \textit{Kepler} metallicities are best fit by the relation,
\begin{equation}
\mathrm{[Fe/H]_{KIC}} = -0.20 + 0.43\mathrm{[Fe/H]_{LAMOST}},
\end{equation}
with the relation most secure in the range $-0.3 < \mathrm{[Fe/H]_{LAMOST}} < 0.4$.
It is thus necessary to interpret any analysis of \textit{Kepler} metallicities with caution. Nevertheless, it should still be meaningful to compare the statistical distribution of candidate metallicities to the rest of the population.

\begin{figure*}
\epsscale{1}
\plotone{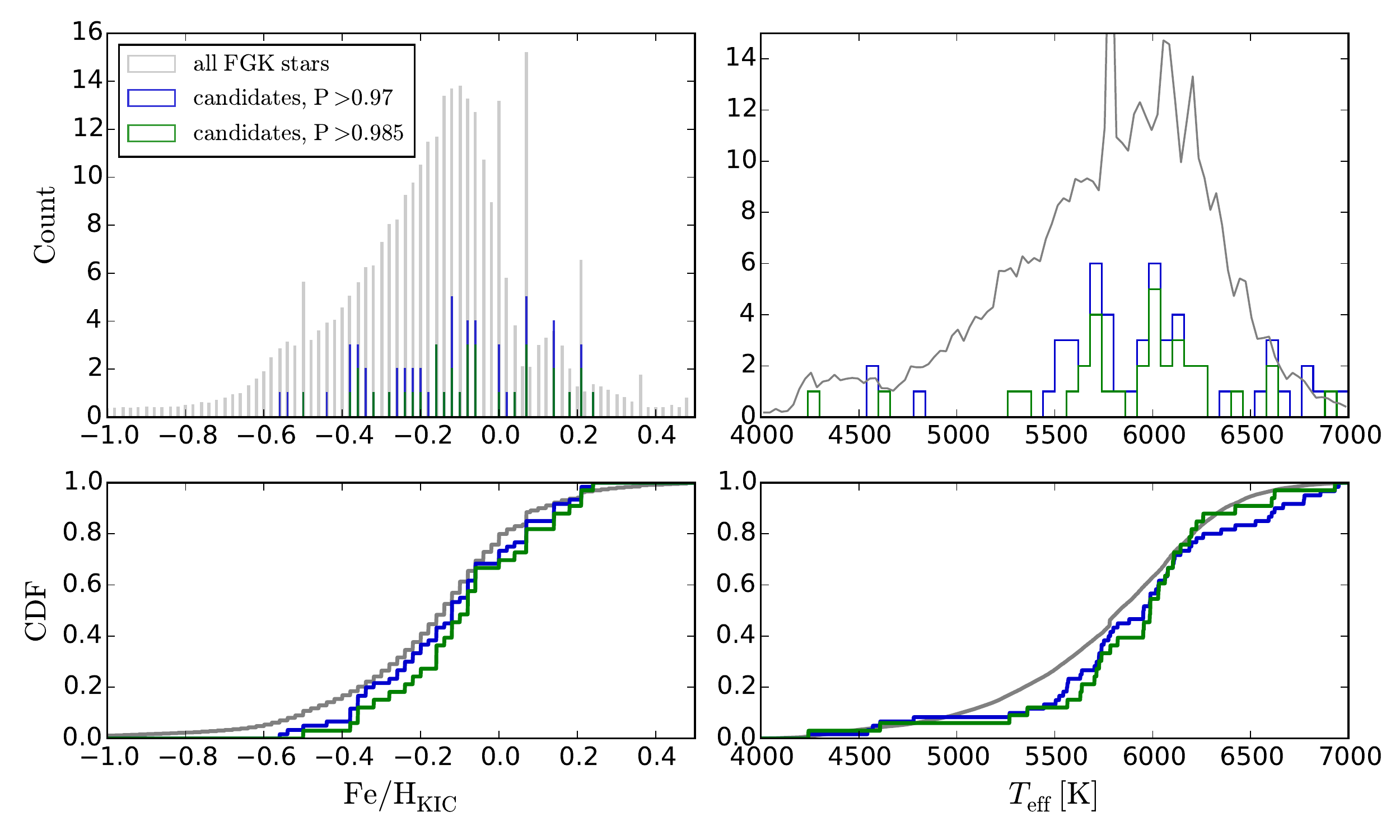}
\caption{The comparison between the \textit{Kepler} stellar catalog metallicities and $T_{\mathrm{eff}}$ for the high probability candidates and for all FGK stars in gray. Candidates with $P>0.97$ are in blue and those with $P>0.985$ in green. The top panels show histograms, where the candidate histograms have been normalized to be displayed on a similar scale as the histograms for all FGK stars. The bottom panels show cumulative distribution functions. It is important to note that the metallicities are systematically underestimated \citep{2014ApJ...789L...3D, 2017arXiv170310400P}. As expected, the metallicity and $T_{\mathrm{eff}}$ distributions for the candidates are systematically high.}
\label{metallicity_Teff_dist}
\end{figure*}

In Figure~\ref{metallicity_Teff_dist}, we display the metallicity histograms and cumulative distribution functions of all FGK stars in gray and of the candidate stars above certain probability thresholds in blue and green. It is clear to see that the metallicities of the candidates are systematically higher than the rest of the FGK stars. The candidates above the $P>0.985$ threshold are shifted further, indicating a smaller amount of false positive contamination as $P$ increases.

An identical analysis may be performed for $T_{\mathrm{eff}}$, since the occurrence of hot Jupiters drops off sharply for $T_{\mathrm{eff}} \lesssim 5000$ K. As shown in Figure~\ref{metallicity_Teff_dist}, $T_{\mathrm{eff}}$ is systematically high for the candidates, which should be expected if the candidates are true giant planets.

\subsection{KOI host stars}
Our analysis has focused on the collection of \textit{Kepler} FGK stars without known planets or planet candidates. Although the search for non-transiting hot Jupiters around \textit{Kepler} transiting planet hosts was the original motivation of this work \citep{2016ApJ...823L...7M}, here we have intentionally excluded them. In KOI systems, the detection of astrometrically-induced transit timing variations for the transiting planet may serve as an additional predictor for the existence of a non-transiting giant planet. These targets are thus best handled separately, and we will be addressing them in future work.

\section{Conclusion}

The prospects of using phase curves to detect non-transiting, Jupiter-mass planets has been an outstanding problem for some time \citep{2011MNRAS.415.3921F, 2014ApJ...795..112P, 2015AJ....150..112S, 2017arXiv170300496S}.  In this paper, we constructed a supervised learning algorithm for the detection of non-transiting \textit{Kepler} planets with optical phase curves.  Our algorithm relies on exploiting the time-dependent properties of planetary phase curves, namely the fact that phase curves are more temporally consistent in their amplitudes and phase compared to other types of light curve variability. 

We first developed a pipeline to identify candidate phase curve signals in \textit{Kepler} light curves. We demonstrated the pipeline's recovery efficiency using light curves containing synthetically injected phase curves. The phase curves recovered by the pipeline exhibited significantly different properties from ``non-phase curves'', that is, signals where the pipeline's detection did not match the injected phase curve. We then developed a logistic regression algorithm in a supervised learning context to classify phase curves and non-phase curves. 
The algorithm performs exceedingly well in its ability to predict phase curves in synthetic datasets and the set of \textit{Kepler} transiting hot Jupiters.

We applied our algorithm to the full set of \textit{Kepler} FGK stars without known planets or planetary candidates and identified 60 high-probability planet candidates. We examined trends in the candidates' albedos and phase offsets and discussed physical explanations for the observed trends. All of our candidates are available for inspection at \href{https://smillholland.github.io/Non-transiting_HJs/}{https://smillholland.github.io/Non-transiting\_HJs/}. 

For more than two decades, the characterization of hot Jupiters has been consistently improved via careful observational study from both the ground and space. Yet many zeroth-order aspects of these bizarre worlds -- including their energetics, their interior structures, and their origins -- remain poorly understood. Aided by the supervised machine learning detection of optical phase curves, a substantial augmentation to the known population of hot Jupiters can potentially be obtained. Planetary candidates identified in this manner can be readily verified with sparsely sampled Doppler velocity observations. These measurements will not only enable the confirmation of new planets, but will also enhance the analysis of the phase curves and aid in the understanding of the planets' atmospheric dynamics. We are particularly optimistic that the techniques outlined here can be refined and extended to forthcoming space-based photometric surveys, including, notably, the TESS and Plato Missions.

\section{Acknowledgements}
We are indebted to the referee, Jon Jenkins, whose expertise in the subject and thorough review of the manuscript helped facilitate significant improvements. S.M. is supported by the National Science Foundation Graduate Research Fellowship Program under Grant Number DGE-1122492. This material is also based upon work supported by the National Aeronautics and Space Administration through the NASA Astrobiology Institute under Cooperative Agreement Notice NNH13ZDA017C issued through the Science Mission Directorate. We acknowledge support from the NASA Astrobiology Institute through a cooperative agreement between NASA Ames Research Center and Yale University. This research has made use of the NASA Exoplanet Archive, which is operated by the California Institute of Technology, under contract with the National Aeronautics and Space Administration under the Exoplanet Exploration Program. This work has also made use of the Exoplanet Orbit Database
and the Exoplanet Data Explorer at \href{exoplanets.org}{exoplanets.org}.

\bibliographystyle{apj}
\bibliography{Systematic_NonTransiting_HJ_Search}

\end{document}